\documentclass[aps,prd,eqsecnum,superscriptaddress]{revtex4-1}

\usepackage{amsfonts}
\usepackage{amsmath}
\usepackage{amssymb}
\usepackage{graphicx}
\usepackage{epsfig}
\usepackage{bm}
\usepackage{mathrsfs}
\usepackage{color}
\usepackage{hyperref}

\begin{document}

\def\bM{ {\bf M} }
\def\bP{ {\bf P} }
\def\br{ {\bf r} }
\def\bu{ {\bf u} }
\def\bX{ {\bf X} }
\def\bx{ {\bf x} }
\def\by{ {\bf y} }
\def\bz{ {\bf z} }
\def\bZ{ {\bf Z} }

\def\dz{ {\dot{z}} }
\def\dq{ {\dot{q}} }
\def\dQ{ {\dot{Q}} }
\def\dzeta{ {\dot{\zeta}} }

\def\xret{ x^0 - | \bx | }
\def\Hhat{ H_{\hat{0} \hat{0} } }

\def\b0{ {\bf 0} }

\def\mbar{ {\bar{m}} }

\def\cP{ {\cal P} }
\def\cQ{ {\cal Q} }
\def\cM{ {\cal M} }
\def\cMout{ { {\cal M}_{out} } }
\def\cMin{ { {\cal M}_{in} } }
\def\cH{ {\cal H} }
\def\cS{ {\cal S} }

\def\th{ \theta }
\def\vf{ \varphi }

\def\intout{ \int_\cMout {\hskip-0.2in} }
\def\intin{ \int_\cMin {\hskip-0.175in} }

\def\lsim{\mathrel{\rlap{\lower3pt\hbox{\hskip1pt$\sim$}}
    \raise1pt\hbox{$<$}}}                
\def\gsim{\mathrel{\rlap{\lower3pt\hbox{\hskip1pt$\sim$}}
    \raise1pt\hbox{$>$}}}  		  

\def\coordeq{ \, \mathrel{ \rlap{\hbox{\hskip-2.5pt$=$} }
    \raise4pt\hbox{$\cdot$}} \, }                

\newcommand{\BL}[1]{{\color{red}$\blacksquare$~\textsf{[xxx BL: #1]}}}
\newcommand{\RB}[1]{{\color{blue}$\blacksquare$~\textsf{[xxx RB: #1]}}}
\newcommand{\CG}[1]{{\color{magenta}$\blacksquare$~\textsf{[xxx CG: #1]}}}

\newcommand{\refme}{{\color{red} [REFME]}}

\title{Theory of optomechanics:  Oscillator-field model of moving mirrors}

\author{Chad R. Galley}
\affiliation{Jet Propulsion Laboratory, California Institute of Technology, Pasadena, CA 91125, USA}
\affiliation{Theoretical Astrophysics, California Institute of Technology, Pasadena, CA 91106, USA}
\author{Ryan O. Behunin}
\affiliation{Center for Nonlinear Studies and Los Alamos National Laboratory, Theoretical Division, Los Alamos, New Mexico 87545, USA}
\author{B. L. Hu}
\affiliation{Joint Quantum Institute and Maryland Center for Fundamental Physics,\\
University of Maryland, College Park, Maryland 20742, USA}
\affiliation{Institute for Advanced Study and Department of Physics, \\ Hong Kong University of Science and Technology, Clear Water Bay, Kowloon, Hong Kong, China}

\date{\today}

\begin{abstract}
In this paper we present a model for the kinematics and dynamics of optomechanics \cite{MarGir} which describe the coupling between an optical field, here modeled by a massless scalar field, and the internal (e.g., determining its reflectivity) and mechanical (e.g., displacement) degrees of freedom of a moveable mirror. As opposed to implementing boundary conditions on the field we highlight the internal dynamics of the mirror which provides added flexibility to describe a variety of setups relevant to current experiments. The inclusion of the internal degrees of freedom in this model allows for a variety of optical activities of mirrors from those exhibiting broadband reflective properties to the cases where reflection is suppressed except for a narrow band centered around the characteristic frequency associated with the mirror's internal dynamics.  After establishing the model and the reflective properties of the mirror we show how appropriate parameter choices lead to useful optomechanical models such as the well known Barton-Calogeracos model \cite{BC} and the important yet lesser explored nonlinear models (e.g., $Nx$ coupling) for small photon numbers $N$, which present models based on side-band approximations \cite{Kimble} cannot cope with. As a simple illustrative application we consider classical radiation pressure cooling with this model. To expound its theoretical structure and physical meanings we connect our model to field-theoretical models using auxiliary fields and the ubiquitous Brownian motion model of quantum open systems. Finally we describe the range of applications of this model, from a full quantum mechanical treatment of radiation pressure cooling, quantum entanglement between macroscopic mirrors, to the backreaction of Hawking radiation on black hole evaporation in a moving mirror analog.
\end{abstract}

\maketitle

\section{Introduction}

Optomechanics deals with the interaction of light with mechanical systems. (For an introductory review see e.g., \cite{MarGir} and references therein.) Though old in name it is relatively new in content -- optomechanics has a history at least as old as radiation pressure \cite{Caves80}. At the quantum level, optomechanics can be traced at least as far back as Casimir \cite{Casimir}, who showed that there is an attractive force between two conducting plates from the change of ground state energy in the presence of boundary conditions, and to Casimir and Polder \cite{CasPol} who calculated the force on an atom near an ideal mirror.  The last decade has seen intense interest in several areas that are all under the umbrella of optomechanics. To name one such area: the dynamical Casimir effect \cite{dynCas} where a moving  object, be it a moving mirror in vacuum, a contracting gas bubble in a fluid (e.g., sonoluminescence as advocated in \cite{snl}),  a time-varying magnetic flux bias threading a SQUID terminating a coplanar transmission line \cite{Wilson11}, or even the spacetime (see below). 
Optomechanics is of renewed current interest because of at least three new developments.
The first relating to nanotechnology \cite{nano}, where miniature mechanical motion can be transduced or manipulated with high precision by capacitive coupling or optical control, or in nano-scale wave guides where radiation pressure effects become important, e.g.
leading to large tailorable photon-phonon couplings which give rise to a large enhancement of stimulated Brillouin scattering  \cite{Rakich12}. The second pertains to quantum information, where information stored in atoms and photons can interface with mechanical devices \cite{Harris,KimbleYeZoller}.
The third pertains to the use of atoms as optical elements \cite{Hetet11}. 

Historically, the gravitation physics community also has explored mirror-field interactions in several ways. For example, cosmological particle creation in the early universe (studied by Parker and Zel'dovich in the 60's-70's \cite{Par69,Zel70}) is a form of the dynamical Casimir effect since it arises from the parametric amplification of vacuum fluctuations by the expansion of the universe. Another example was  the use of a uniformly accelerated mirror as an analog model of Hawking-Unruh effects developed by Davies and Fulling \cite{FulDav,DavFul}) \cite{Haw75,Unr76}. It should be noted that these effects are different from cosmological particle creation as both the black hole and the uniformly accelerated detector/mirror have event horizons while the former in general does not (the de Sitter and anti- de Sitter universes being notable exceptions).  Yet another example, Forward following the suggestion by Joseph Weber \cite{ForWeb} proposed using laser interferometers for the detection of gravitational waves, which has since ushered in today's large-scale and international ground-based gravitational wave detection effort \cite{LIGO, VIRGO, GEO600}.  These gravitational wave detectors are probably the best illustration of the reverse function of optomechanics since in this case an impinging gravitational wave displaces the mirrors in the interferometer and the laser beam in the optical arms picks up the corresponding signal.

In terms of practical applications, an optomechanical process that is actively pursued now is mirror cooling by radiation pressure (see e.g., \cite{mircool}).  Optomechanics also provides an excellent means for probing foundational issues in quantum physics. Sample studies include: 1) Reaching beyond the standard quantum limit using superconducting  \cite{Martini} and nanoelectromechanical \cite{Schwab} devices;  2) Schemes for the improvement of signal-to-noise ratio in gravitational wave interferometer detectors  \cite{BraVya,BuonChen,Buonanno02,ChenSQL} based on earlier theoretical work of Braginsky and Khalil \cite{BraKha}, Caves \cite{Caves80,Caves81}, Unruh \cite{Unruh82}, Kimble, Thorne, et al.~\cite{Kimble};  3) Quantum superposition and entanglement of macroscopic objects such as between a mirror and the field \cite{Marshall} and between two mirrors \cite{Ent2mir}; and 4) Gravitational decoherence, both in its possible limitation to the precision of  atom interferometry \cite{Reynaud} and as a justification for a modified quantum theory \cite{DiosiPenrose}.

Theoretical development for optomechanics also began quite early, most notably in the classic papers of Moore \cite{Moore}, Fulling and Davies \cite{FulDav}, Jaekel and Reynaud \cite{JaeRey},  Barton and Calogeracos \cite{BC}, Law \cite{Law}, Dodonov and Klimov \cite{Dod}, Schu¡§tzhold, et al.~\cite{Schutz} who took a canonical Hamiltonian approach, and Hu and Matacz \cite{HuMat94}, Golestanian and Kardar \cite{Kardar,KardarRMP}, Wu and Lee \cite{WuHsiangLee}, Fosco, Lombardo and Mazzitelli \cite{FLM} who took a path integral approach.  There is also a lineage of work on relativistically moving mirrors as analog models of the Hawking effect (see e.g., \cite{ForVil82,ObaPar,HaroEliz}).  However, many theoretical aspects remain untouched or were treated loosely (some even erroneously). In view of the momentous recent advances in optomechanics, we find it timely and necessary to  construct  a more solid and complete theoretical framework
of moving mirrors interacting with a quantum field.

Our goal is to come up with models and theories capable of treating all of the problems listed above yet conceptually simple  and theoretically systematic enough to be viable and useful.  Admittedly not a simple task \cite{footnote1}, we will delineate different aspects as we progress. Suffice it to mention that this first series of papers present the basic models and theories of optomechanics both  for a closed (this paper) and open (sequel paper in this series)  \cite{StoEqOM} 
system dynamics of moving mirrors in a quantum field. These models can be used to treat the broad class of problems related to the dynamical Casimir effect, among other things. The second series includes the back-action of quantum fields on the mirror, which is needed for treating mirror cooling (for earlier work see references in \cite{BGH_mircool}), the results therein could be applied to the related topics of quantum friction \cite{Pendry97,Volokitin99,Dedkov02,Philbin09}
and vacuum viscosity \cite{Zel70,Hu80,KardarRMP}. A third series will focus on basic issues in quantum information, making use of the stochastic equations derived in a following paper for moving mirrors interacting with a quantum field, specifically on quantum decoherence, superposition, and entanglement of mirrors and field. A different vein using similar techniques but staged in curved spacetimes is the moving mirror analog of backreaction of Hawking radiation on the evolution of a black hole (for background see references in e.g., \cite{BGH_bhbkrn}).

In this first paper we tend to the first order of business, namely, that of developing a useful microscopic model for any number of mirrors interacting with a field.  We consider  a massless scalar field in one spatial dimension for simplicity. The more realistic electromagnetic field in three spatial dimensions can be treated with a slight modification in the form of mirror-field coupling, known as the minimal coupling (see the appendix of \cite{RHA}).
In most prior considerations for the primary functions of a mirror its \emph{reflective properties} (say, by the AMO community) and the \emph{boundary conditions} it imposes on the ambient field (say, by the field theory community), i.e., its amplitude has to vanish at the location of the mirror, are considered in a disjoint manner.  The advantage of the present model is that it avoids the necessity for considering boundary conditions (e.g., {\it a la} Fulling and Davies). Only upon elimination of the explicit dependence of the internal degrees of freedom of the mirror would the field equations require careful attention to boundary conditions. We then consider the kinematics of \emph{mirror motion}, which also has an effect on the field. For example, the motion can parametrically amplify the field modes, including its vacuum fluctuations, which results in particle creation (in the field theory language) or ``motion-induced''/``acceleration'' radiation (in the atom-optics language).

From practical experience physical mirrors have surfaces possessing ``light'' (as opposed to ``heavy'') degrees of freedom that interact with externally incident radiation in such a way as to maintain the appropriate boundary conditions that depend on the material composition of the mirror. Physical mirrors are transparent to sufficiently high frequency components of the field because the mirror's internal degrees of freedom are not energetic enough to (strongly) couple to field modes with arbitrarily high frequencies. For field modes with frequencies far below this cut-off frequency, known as the plasma frequency, the mirror becomes nearly perfectly reflecting.

In this paper, we treat the mirror motion as that of a particle with mass $M$ and corresponding to the center of mass 
of the mirror. To account for the mirror's reflectivity, we model the mirror's ``light'' internal degree of freedom  as a simple harmonic oscillator with mass $m$ and natural frequency $\Omega$. This internal variable  $q(t)$ is taken to couple  bilinearly to the  massless scalar field at the mirror's location. Because this model involves the mutual interaction of the internal oscillator, the field, and the
center of mass motion of the  mirror we shall call this model an {\it mirror-oscillator-field} (MOF) model for optomechanical applications. Further details and properties of the MOF model are given in Section \ref{sec:MOFatrest}.

In Section \ref{sec:MOFatrest} we demonstrate the mirror's ability to reflect and transmit incident radiation and to \emph{perfectly} reflect or transmit radiation upon judicious choices for the parameter values of the internal oscillator. 
We also compare our model with two commonly used models/descriptions for mirrors: 1) The model of Barton and Calogeracos (BC) \cite{BC} (described in  Section \ref{sec:relationtoBCaux}) for partially transmitting mirrors; and  2) The auxiliary field approach of Golestanian and Kardar \cite{Kardar, KardarRMP}. We also show that our  model extends the BC model to  nonadiabatic regimes of the internal oscillator dynamics.
In Section \ref{sec:onemovingMOF} we  turn our attention to  a moving mirror by extending our model  to allow for arbitrary motion, relativistic or non-relativistic.
In Section \ref{sec:multiplemovingMOF} we describe the  MOF model for multiple moving mirrors and focus our attention on how our model appropriately describes multiple reflections and transmissions of  radiation incident on a cavity. Hence, the MOF model is also applicable to multiple-mirror systems in general and to a cavity, in particular, which should be useful for laboratory related studies.
In Section \ref{sec:cooling} we apply the MOF model to describe (classical) mirror cooling by radiation pressure and indicate the role of the mirror's internal oscillator.
In Section \ref{sec:nxcoupling} we show how the bilinear coupling in the MOF model relates to the phenomenological model of moving mirrors wherein the radiation pressure acts on the mirror through the number of incident photons  times the position of the mirror (which we refer to as an $Nx$-type coupling).
In Section \ref{sec:qbm} we show how the MOF model of $N$ moving mirrors is related to models of quantum Brownian motion (QBM) involving $N$ harmonic oscillators coupled to a bath of oscillators. The available and exact master equations for the latter model will facilitate, among  other things, our later studies of entanglement between two mirrors, a prototype problem in macroscopic quantum phenomena as described above.
Finally, in Section \ref{sec:summary} we summarize our findings and mention further work in progress toward the construction of a more complete theory of optomechanics.

\section{A mirror at rest modeled by a bilinear oscillator-field coupling}
\label{sec:MOFatrest}

In this section we introduce a model for a mirror at rest interacting with a scalar field. Our system consists of a mirror with mass $M$ that we treat as being point-like so that, when allowed to move, its trajectory is described by coordinates $Z(t)$. The ``light'' degrees of freedom, which are responsible for the reflective function of the mirror, is modeled as an internal oscillator $q(t)$ with mass $m \ll M$ and natural frequency $\Omega$. For brevity, we will refer to this internal mirror oscillator as a {\it mirosc}. Modeling the ``light'' degrees of freedom by a simple harmonic oscillator is functionally similar to the idealization of the internal degrees of freedom of an atom as a ``two-level'' system  when  considering the atom's optical activities (such as spontaneous and stimulated emission) when interacting with a field via a resonant type of coupling \cite{AH00}  from a ``harmonic atom'' coupling with a bosonic bath  when multiple level activities become important.  Lastly, we take the {\it mirosc} to couple to the external (possibly quantum) scalar field $\Phi(t,x)$ in a manner that is linear in both quantities (i.e., bilinearly coupled). Taken together, we will refer to this model categorically as an {\it mirror-oscillator-field} (MOF) model for optomechanical applications. Different oscillator-field couplings in this model give rise to different models familiar in optics. However, we will always be considering a bilinear coupling in this paper.

We shall show below how this model can describe, with appropriate choices of parameters, a range of perfectly and imperfectly reflecting mirrors. We also show how it relates to the model of Barton and Calogeracos (BC) \cite{BC} used in the quantum optics community and  to the auxiliary field model used more in the field-theory community \cite{Kardar, KardarRMP}.

\subsection{Reflectivity of a mirror modeled by a bilinear oscillator-field coupling}

To demonstrate that the MOF model described above actually possesses the ability to reflect incoming modes it is sufficient to put the mirror at rest at the origin so that the action is given by
\begin{eqnarray}
	S [ \Phi, q ] = \frac{1}{2} \int d^2x \, \partial_\alpha \Phi \partial^\alpha \Phi + \frac{ m }{ 2} \int dt \left( \dot{q}^2 - \Omega^2 q^2 \right)  + \lambda \int dt \, q(t) \Phi (t, 0)
\label{bilinear1}
\end{eqnarray}
where $\eta_{\alpha \beta} = {\rm diag} (1, -1)$  is the metric of 1+1 dimensional Minkowski  space-time. In units where $\hbar = c = 1$ the coupling constant $\lambda$ has dimensions of (mass)${}^{-2}$ = (length)${}^2$.
The equations of motion are obtained by varying (\ref{bilinear1}) in the usual way,
\begin{eqnarray}
	\partial_\alpha \partial^\alpha \Phi = \partial_t^2 \Phi - \partial _x^2 \Phi &=& \lambda q(t) \delta (x) \label{field0} \\
	m \ddot{q} (t) + m \Omega^2 q (t) &=& \lambda \Phi (t, 0) .  \label{oscillator0}
\end{eqnarray}
Let a plane wave with frequency $\omega$ be incident on the mirror from the left ($L$) so that the field is given by
\begin{eqnarray}
	\Phi_{\omega L} = e^{-i \omega t} \left[ \theta (-x) \left( e^{i \omega x} + R (\omega) e^{-i \omega x} \right) + \theta (x) T (\omega) e^{i \omega x} \right]
\end{eqnarray}
where $R(\omega)$ and $T(\omega)$ are the frequency-dependent reflection and transmission coefficients, respectively. For the steady-state evolution of the mirosc-field system we can take $q$ to oscillate with the same frequency as the incident radiation so that
\begin{eqnarray}
	q(t) = A e^{-i \omega t}
\end{eqnarray}
where the amplitude $A$ is determined from (\ref{oscillator0}) to be
\begin{eqnarray}
	A = \frac{ \lambda }{ \Omega^2 - \omega^2 } \, \frac{T(\omega)}{m}  .
\end{eqnarray}
The field is continuous at the location of the mirror, $\Phi_{\omega L} (t,0^+) = \Phi_{\omega L} (t,0 ^-)$, and the discontinuity of the spatial derivative is found by integrating (\ref{field0}) over a vanishingly small interval encompassing the mirror's position,
\begin{eqnarray}
	- \partial_x \Phi_{\omega L} (t,0^+) + \partial_x \Phi_{\omega L} (t,0^-) = \lambda q(t)  .
\end{eqnarray}
Using these conditions and the equations of motion in (\ref{field0}) and (\ref{oscillator0}), the reflection and transmission coefficients are given in terms of the three mirosc parameters $(m, \Omega, \lambda)$ by
\begin{eqnarray}
	R (\omega) &=& - \frac{ i \lambda^2 }{ 2m \omega (\Omega^2 - \omega^2) + i \lambda^2 } \label{Rbilin}  \\
	T (\omega) &=& \frac{ 2m \omega ( \Omega^2 - \omega^2) }{ 2m \omega (\Omega^2 - \omega^2) + i \lambda^2 }   \label{Tbilin}  .
\end{eqnarray}

There are three ways that the particle can perfectly reflect incident radiation: 1) In the strong oscillator-field coupling limit, $\lambda \to \infty$; 2) When the oscillator is resonantly excited by the monochromatic radiation, $\Omega = \omega$, (independently of the values of $\lambda$ and $m$); and 3) In the limit that the mass of the mirosc  vanishes, $m \to 0$. In all three cases, the reflection and transmission coefficients are $R (\omega) = -1$ and $T (\omega) = 0$, respectively.

Likewise, perfect transmission can be attained in three manners: 1) In the limit of vanishingly small oscillator-field coupling, $\lambda \to 0$; 2) When the mirosc frequency is arbitrarily large, $\Omega \to \infty$; and 3) When the  mirosc mass is arbitrarily large, $m \to \infty$. In all three cases, $R (\omega) = 0$ and $T (\omega) = 1$.

\begin{figure}
\begin{center}
\includegraphics[width=0.66 \textwidth]{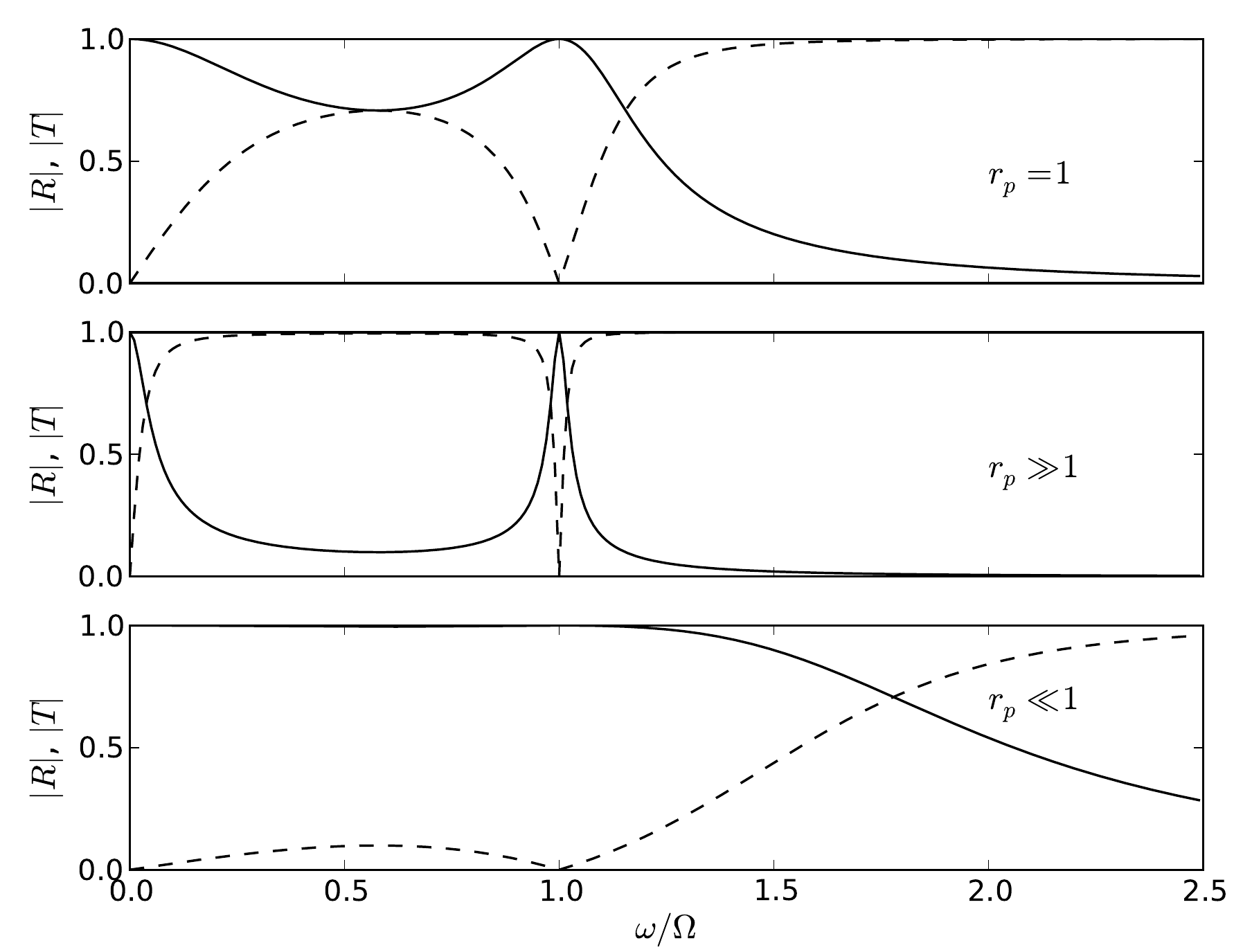}
\caption{Complex norm of the reflectivity (solid) and transmissivity (dashed) for the case when the minimum in $|R(\omega)|^2$ is 50\% (top) and when the oscillator reflects incident modes in a narrow (middle) and a broad bandwidth (bottom).}
\label{fig:reflection}
\end{center}
\end{figure}

The reflection and transmission coefficients possess interesting features that depend on the parameters of the theory $(m, \Omega, \lambda)$. The squared complex norm of the reflection coefficient from (\ref{Rbilin}) is
\begin{eqnarray}
	| R (y) | ^2 = \frac{ 1 }{ 1 + \left[ \displaystyle\frac{ 2m \Omega^3 }{ \lambda^2 } y (1-y^2) \right] ^2 }
\label{Rnormsq}
\end{eqnarray}
where we have introduced the dimensionless quantity $y = \omega / \Omega$.
To characterize the dependence of the reflection coefficient on the frequency $\omega$ of the incident field mode we observe that the local minima and maxima occur for $y_{\rm min} = 1/ \sqrt{3}$ and $y_{\rm max} = \{ 0, 1 \}$, respectively. The maximum at $\omega = 0$ is an artifact of the monopole coupling between the field and the mirosc. For a dipole coupling the reflection coefficient vanishes at $\omega = 0$. The reflection coefficient equals $1$ at both maxima and the value of $|R|^2$ at $y_{\rm min}$ is
\begin{eqnarray}
	| R (y_{\rm min} ) |^2 = \frac{1}{ 1 + r_p^2 }
\end{eqnarray}
where we define the {\it plasma frequency} $\Omega_p$ for the partially transmitting mirror to be
\begin{eqnarray}
	\Omega_p \equiv \frac{ 3^{3/2} \lambda^2 }{ 4 m \Omega^2}
\end{eqnarray}
and the index $r_p \equiv \Omega/ \Omega_p = 4m \Omega^3 / ( 3^{3/2} \lambda^2)$.

We can use this minimum in the reflected radiation to indicate when the two maxima of $|R|^2$ are sufficiently separated and distinguishable. While this is subjective we take $| R (y_{\rm min} ) |^2  = 1/2$ to be our defining requirement, which fixes $r_p = 1$. The implication is that if $r_p \gg 1$ then the reflection coefficient is sharply peaked about $\omega = \Omega$. Under this condition, the parameters of the mirosc can be tuned to selectively reflect incident radiation in a narrow bandwidth centered on $\omega = \Omega$, which occurs when the  mirosc is resonantly excited (or very nearly so) by the incident field.
Whenever the mirosc parameters are such that $r_p \ll 1$ then the local minimum at $y_{\rm min} = 1/\sqrt{3}$ is close to $1$ and the mirosc reflects modes over a broader frequency bandwidth. Furthermore, the mirror will reflect more than half of the incident radiation so long as the frequency of the field is less than $\sim \Omega$ to a good approximation. Hence, if the mirosc mass is made smaller or the oscillator-field coupling constant $\lambda$ is made larger then more modes  will be reflected more strongly by the mirror. Fig.~(\ref{fig:reflection}) shows the basic features of the mirror's scattering properties studied in this section and contains plots of the reflection coefficient $|R(y)|^2$ for $r_p$ equal to, much larger, and much smaller than $1$.

\subsection{Relation to common mirror models and approaches}
\label{sec:relationtoBCaux}

The bilinearly-coupled oscillator-field model introduced above possesses  interesting physical limits that  relate to two well-known and commonly used mirror models. The first model arises when the mirosc evolves adiabatically with the field and gives rise to the model of Barton and Calogeracos (BC) for a partially transmitting mirror. The second model arises when the mass of the mirosc becomes arbitrarily small, in which case the mirosc serves the role of an auxiliary field that relates to the path integral approaches of \cite{Kardar, KardarRMP, FLM},  which describe a quantum field interacting with a perfectly reflecting mirror(s). In Section \ref{sec:nxcoupling} we also relate the MOF formulation of optomechanics to a commonly used model which describes  the effects of radiation pressure by invoking a phenomenological coupling between the number of photons impinging the mirror and the mirror's position.

\subsubsection{Barton-Calogeracos model}

The BC model has been used quite often in quantum optics and it is worth summarizing its primary properties before showing how it can be derived from our MOF model.
Much of BC's attention focuses on quantizing the non-relativistic limit of the theory where the mirror velocity is much smaller than $c$. We do not present their results here but refer the reader to the original papers of \cite{BC} for further details.

The action for a mirror at rest in the BC model is, in 1+1 dimensions of space-time,
\begin{eqnarray}
	S_{\rm BC} [\Phi] = \frac{1}{2} \int d^2x \, \partial_\alpha \Phi \partial^\alpha \Phi - \gamma \int dt \, \Phi^2 (t, x=0)
\end{eqnarray}
where $\gamma$ is related to the plasma frequency of the mirror \cite{BC}. Extremizing the action gives the  equations of motion
\begin{eqnarray}
	\partial_\alpha \partial^\alpha \Phi = 2 \gamma \delta (x ) \Phi (t, 0)   \label{BC_eom}
\end{eqnarray}
The reflection and transmission of a normal mode of the field incident on the mirror from the left ($x<0$) is
 \begin{eqnarray}
 	\Phi_{\omega L} = e^{-i \omega t} \left[ \theta (-x) \left( e^{i \omega x} + R (\omega) e^{-i \omega x} \right) + \theta (x) T (\omega) e^{i \omega x} \right]  \label{BCnormal0}
 \end{eqnarray}
 where $R (\omega)$ and $T (\omega)$ are the frequency-dependent reflection and transmission coefficients, reflectively, with the property that $|R|^2 + |T|^2 = 1$.

 We demand that the field be continuous across the mirror $\Phi_{\omega L} (t,0^+) = \Phi_{\omega L}(t,0^-)$ and that its derivative satisfy
\begin{eqnarray}
	- \partial_x \Phi_{\omega L} (t,0^+) + \partial_x \Phi_{\omega L} (t,0^-) = 2 \gamma \Phi_{\omega L} (t,0 ).  \label{BC_jump}
\end{eqnarray}
 This jump condition follows from integrating the field equations across the mirror's position at $x=0$. Together with the field equation these conditions imply that
 \begin{eqnarray}
 	R (\omega) &=& - \frac{ i \gamma }{ \omega + i \gamma } \\
	T (\omega) &=& \frac{ \omega }{ \omega + i \gamma }   .
 \end{eqnarray}
As the parameter $\gamma$ becomes arbitrarily large we see that the reflection becomes perfect and the incoming phase of the field changes by $\pi$ radians
\begin{eqnarray}
	\lim_{\gamma \to \infty} R (\omega) = -1
\end{eqnarray}
The ability of the BC model to reproduce perfect and imperfect reflection comes from using the quadratic interaction $\Phi^2 (t, 0)$. With this specific coupling to the mirror the jump condition across the origin (\ref{BC_jump}) is linear in $\Phi$ at the mirror, which is vital for obtaining the normal mode in (\ref{BCnormal0}).

The MOF model in (\ref{bilinear1}) can be related, under appropriate conditions, to the BC model. Observe from (\ref{oscillator0}) that if $q(t)$ evolves adiabatically with time,
\begin{eqnarray}
	\left| \frac{ \ddot{q} }{ \Omega^2 q } \right| \ll 1  ,
\end{eqnarray}
then the  mirosc follows the time-development of the field at the mirror's position
\begin{eqnarray}
	q (t ) \approx \frac{ \lambda }{ m \Omega^2 } \, \Phi (t,0)
\label{BClimit1}
\end{eqnarray}
Substituting this approximation for the oscillator variable into the scalar field equation (\ref{field0}) gives
\begin{eqnarray}
	\partial_\alpha \partial^\alpha \Phi \approx \left( \frac{ \lambda^2 }{ m \Omega^2 } \right) \delta (x) \Phi(t, 0)  .
\end{eqnarray}
Comparing with (\ref{BC_eom}) we recover the model of BC by identifying $\gamma$ with the parameters of the mirosc and hence to the plasma frequency of the MOF model
\begin{eqnarray}
	\gamma = \frac{ \lambda^2 }{ 2 m \Omega^2 } =  \frac{2}{ 3^{3/2}} \Omega_p  \label{BCtobilin}  .
\end{eqnarray}
Therefore, in the limit that the mirosc changes adiabatically the MOF model yields the BC model.

An equivalent way of connecting to the BC model is to take the mass of the mirosc to zero, $m \to 0$ but keep the quantity $m \Omega^2 \equiv \kappa$ constant in this limit, which requires the mirosc natural frequency to approach infinity, $\Omega \to \infty$. In this limit, the mirosc also follows the time-development of the field
\begin{align}
	q(t) \to \frac{ \lambda }{\kappa } \, \Phi (t, 0).
\label{BClimit2}
\end{align}
The identification with the BC model then follows the same steps as in the previous paragraph and, in particular, one finds that $\gamma = \lambda^2 /( 2 \kappa )$.
It is worth pointing out that the massless limit $m \to 0$ here does not imply that the mirror is perfectly reflecting as in the previous section. This is because of the additional requirement that $m \Omega^2 = \kappa$ remain constant. In fact, the reflection coefficient  (\ref{Rbilin}) in this limit becomes
\begin{align}
	R(\omega)  \to - \frac{ i \lambda^2 }{ 2 \omega \kappa + i \lambda^2 }
\end{align}
and the mirror becomes perfectly reflecting when $\lambda \to \infty$.

Through the identification in (\ref{BCtobilin}) we may attach heuristic physical interpretations to $m$, $\Omega$ (or $\kappa$) and $\lambda$. In \cite{BC1}, Barton and Calogeracos observe that their model is equivalent to a jellium sheet of zero width, i.e., a surface of vanishing thickness having a surface current density generated by the motion of small charge elements with charge density $n_s$. If these elements have charge $n_s e$ per unit area and mass $n_s m_e$ per unit area then BC find
\begin{eqnarray}
	\gamma = \frac{ 2 \pi n_s e^2 }{ m_e }  .
\end{eqnarray}
Identifying these microscopic variables to those in our MOF model via (\ref{BCtobilin}) gives the following relationship
\begin{eqnarray}
	\frac{ 4\pi n_s e^2}{ m_e} = \frac{ \lambda^2}{ \kappa }  .
\end{eqnarray}
This suggests identifying the mirosc field coupling as a charge per unit area, $\lambda \to n_s e$, and $\kappa$ as a mass per unit area, $\kappa \to n_s m_e / (4\pi)$. That is, $\lambda$ can be viewed as a surface charge density and $\kappa = m \Omega^2$ as a surface mass density. This interpretation may be useful for developing a similar MOF model for a mirror in 3+1 dimensions.

\subsubsection{Models using auxiliary fields}

The MOF model reduces to another well-known description of mirrors if we take the limit $m \to 0$. In this limit our model describes a perfectly reflecting mirror, as discussed earlier, and the action (\ref{bilinear1}) becomes
\begin{eqnarray}
\label{Aux-Action}
	\lim _{m \to 0} S [\Phi, q] &=& \frac{1}{2} \int d^2x \, \partial_\alpha \Phi \partial^\alpha \Phi + \lambda \int dt \, q(t) \Phi (t, 0).
\end{eqnarray}
The key point is that the mirosc possesses no dynamics in this limit. Thus,
the quantity $\psi(t) \equiv \lambda q(t)$ possesses no dynamics of its own and can be regarded as an {\it auxiliary field}.

In the path-integral formulation of the quantum theory, the massless mirosc limit gives rise to the following generating functional \cite{footnote}
\begin{align}
	\lim_{m \to 0} Z [J] = \int {\cal D} \Phi \int {\cal D} \psi \, \exp \bigg\{ \frac{ i}{2} \int d^2x \, \partial_\alpha \Phi \partial^\alpha \Phi + i  \int dt \, \psi(t) \Phi(t,0) + i \int d^2x \, J \Phi \bigg\}
\label{eq:Zmassless1}
\end{align}
Then, noting that the path integral over $\psi(t)$ is just the Fourier representation of the Dirac delta functional,
\begin{align}
	\int {\cal D}\psi \, \exp \bigg\{ i \int dt \, \psi(t) \Phi(t, 0) \bigg\} = \delta \big[ \Phi (t,0) \big]
\end{align}
it follows that the generating functional
\begin{align}
	\lim_{m \to 0} Z [J] = \int {\cal D} \Phi \,\delta \big[ \Phi(t,0) \big]  \exp \bigg\{ \frac{ i}{2} \int d^2x \, \partial_\alpha \Phi \partial^\alpha \Phi + i \int d^2x \, J \Phi \bigg\}
\label{eq:Zmassless2}
\end{align}
describes a quantum scalar field constrained to vanish at the location of the mirror (only those field configurations that vanish at $x=0$ will contribute to the path integral). The vanishing of the field at the location of the mirror is equivalent to the perfect reflection of an incident field \cite{Kardar, KardarRMP}.

\begin{figure}
	\includegraphics[height=7cm]{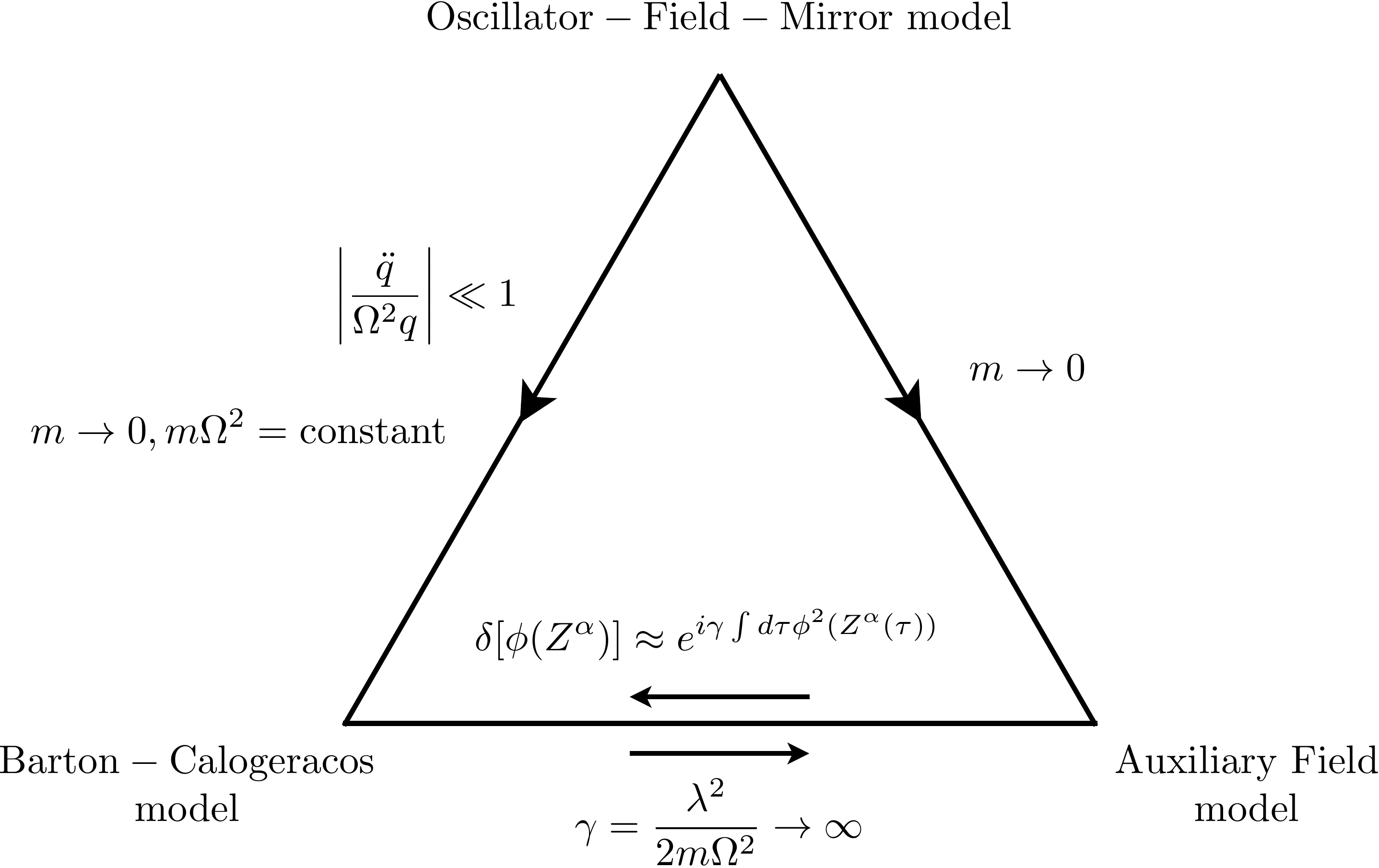}
\caption{Relationships among our bilinearly-coupled oscillator-field-mirror model of a partially transmitting mirror, the Barton-Calogeracos model of a partially transmitting mirror, and the auxiliary field approach for a perfectly reflecting mirror.}
	\label{relationships}
\end{figure}

Our bilinearly-coupled MOF model (\ref{bilinear1}) has successfully reproduced two models describing the interactions of a field with a mirror at rest: 1) The partially transmitting BC mirror model  when $q(t)$ evolves adiabatically;  and 2) an auxiliary field approach that enforces the field to vanish at  the mirror when the mass of the mirosc is vanishingly small. In turn, these two models can be related to each other. Specifically, noting that the delta functional above can be approximated by a narrow Gaussian it follows that (\ref{eq:Zmassless2}) becomes
\begin{align}
	\lim _{m \to 0 } Z[ J] \approx \int {\cal D} \Phi \, \exp \bigg\{ \frac{ i}{2} \int d^2x \, \partial_\alpha \Phi \partial^\alpha \Phi - i \gamma \int d^2 x \, \Phi^2 (t, 0) + i \int d^2x \, J \Phi \bigg\}
\end{align}
which is increasingly more accurate for larger values of $\gamma$.
Hence, BC falls out from the generating functional approach if we smear the delta functional constraint that enforces the field to vanish on the surface of the mirror. Likewise, using the action for the BC model in the generating functional formalism gives the perfect reflection limit when $\gamma \to \infty$. See Fig.~(\ref{relationships}) for the relationships among these theories.

\section{A moving mirror in the MOF model}
\label{sec:onemovingMOF}

As pointed out in the Introduction the physics is quite different in the two cases when the mirror is moving relativistically compared to the case when it is moving slowly. The former relates to cosmological particle creation and radiation emitted from black holes or in uniformly accelerated detectors in the Hawking-Unruh effects while the latter is closer to accessible laboratory situations such as mirror movements caused by the passing of gravitational waves in  interferometer detectors and  mirror cooling from the field's back-action in the form of radiative pressure and quantum friction. The MOF model presented here provides a unified framework for treating both, albeit very different, situations.  
For cases when the mirror motion is prescribed such as
coplanar waveguides terminated by a SQUID  \cite{Nation09,Wilson11},
or when the mirror possesses non-trivial reflective properties \cite{Hetet11}
our model can meet the needs of current experiments
by providing a rich set of reflective properties
and a tractable formalism capable of providing analytical insight.
For systems where the mirror motion
is dynamically determined by the mutual interaction of the mirror's center
of mass, it's internal motion, and the field
our model provides a computational ease.
This simplification  results from the fact that  boundary conditions are not
imposed on the field from the outset but determined by a self-consistent
elimination of the mirror's internal motion. This facilitates the derivation
of equations of motion for the mirror's mean position which will be adopted
in Sec.~\ref{sec:cooling} to describe classical radiation pressure cooling, and in later
papers in this series to provide a fully  quantum mechanical treatment of mirror cooling and
the mirror-analog of the black hole back-reaction.

\subsection{Action formulation}

Allowing the mirror to move requires the addition of an extra term describing its motion along the worldline $Z^\mu (\lambda)$ where $\lambda$ is an affine parameter and $\mu = 0, 1$. The physics must remain invariant under any reparameterization of the mirror's worldline $\lambda \to \lambda ( \bar{\lambda})$, which requires modifying the action (\ref{bilinear1}) for a static mirror in the following way 
\begin{align}
	S [ \Phi, q, Z^\mu] = {} & \frac{1}{2} \int d^2 x \, \partial_\alpha \Phi \partial^\alpha \Phi + \frac{m}{2} \int d\lambda \bigg( \frac{\dot{q}^2 }{ \sqrt{U^\alpha U_\alpha} } - \Omega^2 q^2 \sqrt{ U^\alpha U_\alpha} \bigg) \nonumber \\
		& - M \int d\lambda \, \sqrt{U^\alpha U_\alpha} + \lambda \int d\lambda \, \sqrt{U^\alpha U_\alpha} \, q(\lambda) \Phi \big(  Z^\mu(\lambda) \big)
\label{bilinear2}
\end{align}
where an overdot denotes differentiation with respect to the worldline parameter, $U^\mu(\lambda)=\dot{Z}^\mu (\lambda)$ is the 2-velocity of the mirror, $d\lambda \sqrt{U^\alpha U_\alpha} = d\tau$ is the invariant proper time element as measured by an observer on the worldline, and indices with Greek letters are raised and lowered with the Minkowski metric $\eta_{\alpha \beta} = {\rm diag} (1, -1)$.
The field still couples bilinearly to the mirosc via the last term of the action so that the reflective properties studied in the previous section are retained by the model. The corresponding Euler-Lagrange equations of motion are easily found to be
\begin{align}
	\partial_t ^2 \Phi - \partial_x^2 \Phi & = \lambda q(\tau) \delta^2 ( x^\mu - Z^\mu(\tau) )
\label{eq:ELeom1} \\
	m \ddot{q} + m \Omega^2 q & = \lambda \Phi ( Z^\mu(\tau) )
\label{eq:ELeom2} \\
	M_{\rm eff}(\tau) \dot{U}^\mu & = - \lambda q(\tau) \big( \eta^{\mu\nu} - U^\mu U^\nu \big) \partial_\nu \Phi (Z^\mu(\tau))
\label{eq:ELeom3}
\end{align}
where we conveniently have chosen to parameterize the worldline by the proper time $\tau$ at this point since then $U^\alpha U_\alpha = 1$ and $U^\alpha \dot{U}_\alpha = 0$, which help simplify the expressions. The quantity $M_{\rm eff}(\tau)$ in (\ref{eq:ELeom3}) is an effective mass for the mirror and is given by
\begin{align}
	M_{\rm eff} (\tau) \equiv M + \frac{1}{2} m \dot{q}^2(\tau) + \frac{1}{2} m \Omega^2 q^2(\tau) - \lambda q(\tau) \Phi (Z^\mu(\tau) ) .
\end{align}
Notice that the effective mass has contributions from the rest mass of the mirror ($M$), the energy of the oscillator ($m\dot{q}^2/2 + m \Omega^2 q^2/2$), and the interaction energy of the mirror-oscillator-field system ($-\lambda q \Phi (Z)$). In other words, the effective mass is the rest mass of the mirror plus the total internal energy of the mirosc.

The structure of (\ref{eq:ELeom1}) and (\ref{eq:ELeom3}) is reminiscent of a field coupled to a scalar point charge, which here is played by the time-dependent mirosc amplitude $q(t)$. In $3+1$ dimensions, such a system exhibits a radiation reaction force on the charge proportional to the third time derivative of the particle's position and exhibits the infamous class of unphysical runaway solutions in the absence of any external influences. Below, we show that no such unphysical solutions manifest in our MOF model  here. To show this, we first solve the field equation in (\ref{eq:ELeom1}), which gives
\begin{align}
	\Phi (x^\alpha) = \lambda \int d\tau' \, G(x^\alpha;  Z^\mu(\tau') ) q(\tau')
\label{eq:Phi1}
\end{align}
where we ignore the homogeneous solution and where the retarded Green's function in 1+1 spacetime dimensions is
\begin{align}
	G( x^\alpha; x'^\alpha) = \frac{1}{2} \theta (t-t') \theta \big( \sigma (x^\alpha, x'^\alpha) \big)
\end{align}
where $\sigma$ is half of the squared distance between $x^\alpha$ and $x'^\alpha$ as measured by the straight line (i.e., a geodesic) connecting them, namely,
\begin{align}
	\sigma ( x^\alpha, x'^\alpha ) & = \frac{1}{2} (x^\alpha - x'^\alpha) (x_\alpha - x'_\alpha)
\label{synge1}
\end{align}

The derivative of the field evaluated  on the worldline is then
\begin{align}
	\partial_\nu \Phi (Z^\mu(\tau) ) \equiv {} \big[ \partial_\nu \Phi (x^\alpha) \big]_{x^\alpha =  Z^\alpha (\tau) } =  {} & \frac{\lambda }{ 2} \int d\tau' \, \theta(\tau - \tau')  \partial_\nu \sigma (Z^\mu (\tau), Z^\mu (\tau') ) \delta \big( \sigma ( Z^\mu (\tau), Z^\mu (\tau') ) \big) q(\tau') \\
	& + \frac{ \lambda }{ 2} \int d\tau' \, \partial_\nu (\tau-\tau') \delta (\tau - \tau') \theta \big( \sigma (Z^\mu (\tau) , Z^\mu (\tau') \big) q (\tau')
\label{eq:dPhi1}
\end{align}
From (\ref{synge1}) it follows that since the mirror's worldline is time-like then $\sigma (Z^\mu (\tau), Z^\mu (\tau'))$ is always positive except at $\tau' = \tau$ where it vanishes. Hence, the delta function in the first line of (\ref{eq:dPhi1}) receives a contribution only at coincidence, when $\tau' = \tau$. Likewise, the $\delta (\tau- \tau')$ in the second line of (\ref{eq:dPhi1}) gives support to the integral at coincidence.

To evaluate the integrals in (\ref{eq:dPhi1}) we will need to determine the behavior of $\delta (\sigma)$ and $\partial_\nu (\tau - \tau')$ when $\tau' \approx \tau$. This follows by expanding (\ref{synge1}) around $s \equiv \tau' - \tau$ near zero, giving
\begin{align}
	\sigma (Z^\mu (\tau) , Z^\mu (\tau') ) = \frac{s^2 }{2} - \frac{ s^4}{24} \dot{U}^\alpha(\tau) \dot{U}_\alpha (\tau) + O(s^5)
\label{synge2}
\end{align}
where we have used the identities $U^\alpha U_\alpha = 1$, $U^\alpha \dot{U}_\alpha = 0$, and $U^\alpha \ddot{U}_\alpha = - \dot{U}^\alpha \dot{U}_\alpha$, which are valid in the proper time parameterization of the worldline. Therefore, writing the delta function in (\ref{eq:dPhi1}) as a delta function of $s$ and then expanding (\ref{synge2}) for $s$ near zero gives
\begin{align}
	\delta \big( \sigma (Z^\mu (\tau), Z^\mu (\tau') ) \big) = {} & \frac{ \delta (s) }{ |s| } \bigg( 1 + \frac{s^2}{6}  \dot{U}^\alpha \dot{U}_\alpha + O ( s^3) \bigg).
\end{align}
In addition, the second integral in (\ref{eq:dPhi1}) is proportional to
\begin{align}
	\int ds \, \partial_\nu s \, \delta (s) \theta (\sigma) q( \tau+s) = \frac{1}{2} q(\tau) \big[ \partial_\nu s \big]_{s=0}  .
\end{align}
The important point to note is that the first integral in (\ref{eq:dPhi1})  is potentially divergent. However, we will show now that no divergence actually manifests.

To see this, we observe that (\ref{synge1}) implies $\partial_\nu \sigma (x^\alpha, x'^\alpha) = x_\nu - x'_\nu$, which, when evaluated on the worldline and expanding around $s$ equal zero, yields
\begin{align}
	\partial_\nu \sigma( Z^\mu (\tau), Z^\mu (\tau') ) = - s U_\nu (\tau) - \frac{ s^2}{ 2} \dot{U}_\nu + O(s^3)
\end{align}
Note also that the above equation  implies that $[ \partial_\nu s ] _{s=0} = - U_\nu (\tau)$ since from (\ref{synge2}) it follows that $\partial_\nu \sigma = s \partial_\nu s + O(s^3)$.
The integral in (\ref{eq:dPhi1}) thus becomes
\begin{align}
	\partial_\nu \Phi (Z^\mu( \tau)) = {} & \frac{ \lambda }{ 2} \int_{-\infty}^\infty ds \, \theta (-s) \big( -s U_\nu (\tau) +O(s^2) \big) \big( q(\tau) + O(s) \big) \frac{ \delta (s) }{ |s| } \big( 1 + O(s^2) \big) + \frac{ \lambda}{4} U_\nu (\tau) q(t)  .
\end{align}
Evaluating the integral over $s$ and using $2 \theta(-s) = 1 - {\rm sgn}(s)$ we find that
\begin{align}
	\partial_\nu \Phi (Z^\mu( \tau)) = {} & \frac{ \lambda}{2} U_\nu (\tau) q(t) ,
\end{align}
which is  finite.
In addition, the derivative of the field above, which is proportional to $U_\nu$, is contracted with $\eta^{\mu\nu} - U^\mu U^\nu$ in (\ref{eq:ELeom3}) to get the force on the mirror,  thereby giving {\it zero}.
Hence, the equation of motion for the mirror's worldline from (\ref{eq:ELeom3}) is simply
\begin{align}
	\dot{U}^\mu (\tau) = 0
\end{align}
and the mirror moves inertially. The reason for this trivial dynamics is because the field is not generated by any external sources and because we have ignored the initial configuration of the field (i.e., homogeneous solutions to the field equation (\ref{eq:ELeom1})). Both of these types of sources will impart a non-trivial dynamics for the mirror's motion.

\subsection{Hamiltonian formulation}

Here, we provide a Hamiltonian formulation of the MOF model. To do this, we find it convenient to parameterize the worldline by the coordinate time $t$ wherein the action (\ref{bilinear2}) becomes 
\begin{align}
		S [ \Phi, q, Z] = {} & \frac{1}{2} \int d^2 x \, \partial_\alpha \Phi \partial^\alpha \Phi + \frac{m}{2} \int dt \bigg( \frac{\dot{q}^2 }{ \sqrt{1 - U^2} } - \Omega^2 q^2 \sqrt{ 1 - U^2} \bigg) \nonumber \\
		& - M \int dt \, \sqrt{1 - U^2} + \lambda \int dt \, \sqrt{1 - U^2} \, q(t) \Phi \big(  t, Z(t) \big)
\label{bilinear3}
\end{align}
where $U(t) = \frac{ d Z(t)}{dt}$ and 
from which the Lagrangian is
\begin{align}
	L = {} & \frac{1}{2} \int d x \, \partial_\alpha \Phi \partial^\alpha \Phi + \frac{m}{2}  \bigg( \frac{\dot{q}^2 }{ \sqrt{1 - U^2} } - \Omega^2 q^2 \sqrt{ 1 - U^2} \bigg)  - M  \sqrt{1 - U^2} + \lambda  \sqrt{1 - U^2} \, q(t) \Phi \big(  t, Z(t) \big)  .
\label{lagrangian00}
\end{align}
To derive the Hamiltonian $H$ we first identify the conjugate momenta,
\begin{align}
	\Pi (t, x) & = \frac{ \partial L }{ \partial \dot{\Phi}(t,x) } = \dot{\Phi} (t,x) \\
	p(t) & = \frac{ \partial L }{ \partial \dot{q} (t) } = \frac{ m \dot{q}(t) }{ \sqrt{ 1 - U^2(t)} } \\
	P(t) & = \frac{ \partial L}{ \partial \dot{Z} (t) } = \frac{ M _{\rm eff} (t) U(t) }{ \sqrt{ 1 - U^2(t) } }
\end{align}
where the effective mass in terms of the conjugate momenta is
\begin{align}
	M_{\rm eff} (t) = M + \frac{ p^2(t) }{ 2 m } + \frac{ 1}{2} m \Omega^2 q^2 (t) - \lambda q(t) \Phi (t, Z(t)) .
\end{align}
The Legendre transformation of (\ref{lagrangian00}) yields the Hamiltonian after some algebra
\begin{align}
	H = {} & \frac{ 1}{2} \int dx \, \big( \Pi^2 + (\partial_x \Phi)^2 \big)  + \sqrt{ P^2 + M_{\rm eff}^2 (t) }  .
\label{hamiltonian0}
\end{align}
For completeness, we give Hamilton's equations of motion
\begin{align}
	\dot{\Phi} & = \Pi \\
	\dot{\Pi} & = \Phi'' + \frac{ \lambda q(t) M_{\rm eff} (t)  }{ \sqrt{ P^2 + M_{\rm eff}^2 (t) } } \,  \delta (x - Z(t)) \label{Pi1} \\
	\dot{q} & = \frac{ p}{m} \\
	\dot{p} & = - \frac{  M_{\rm eff} (t)  }{ \sqrt{ P^2 + M^2 _{\rm eff} (t) } } \,\big( m \Omega^2 q - \lambda \Phi (t, Z(t) \big) \\
	\dot{Z} & = \frac{  P }{ \sqrt{ P^2 + M_{\rm eff} ^2 (t) } } \label{Z1} \\
	\dot{P} &  = \frac{ \lambda q(t) M_{\rm eff} (t) }{ \sqrt{ P^2 + M_{\rm eff}^2 (t) } } \, \partial_x \Phi (t, Z(t)) , \label{P1}
\end{align}
which can be shown to be equivalent to the Euler-Lagrange equations in (\ref{eq:ELeom1})-(\ref{eq:ELeom3}). As discussed in the previous section, an external source will be needed to generate non-trivial forces on  the mirror.

Depending on the application, it may be more convenient to work in a reference frame wherein the interaction between the field and the mirror's worldline decouple from each other so that the mirror always remains at rest at the origin. A transformation to such a non-inertial frame is advocated in \cite{BC} and may be useful for canonically quantizing the MOF model.
However, we will not pursue this representation here.

\subsection{A slowly moving mirror in the MOF model}

Under all laboratory conditions to date the speed of the mirror is small compared to $c$ and justifies developing the non-relativistic limit of the mirror-oscillator-field model. For example, it was recently demonstrated that film bulk acoustic resonators (FBARs) \cite{FBAR} as large as $\approx 0.5$mm can be mechanically oscillated up to $3$GHz. The corresponding speed of the FBAR (having a modulation depth of $10^{-8}$) is only $v \approx 4.4$m/s, which is much smaller than $c$. Thus, for laboratory applications, the non-relativistic limit of the MOF action in (\ref{bilinear2}) is entirely appropriate. 

The relativistic Lagrangian (\ref{lagrangian00}) expanded in powers of $\dot{Z} \ll 1$ and retaining the lowest order contributions in the velocity yields
\begin{eqnarray}
	L = \frac{ 1}{2} \int dx \, \partial_\alpha \Phi \partial^\alpha \Phi + \frac{1}{2} m \dot{q}^2 - \frac{1}{2} m \Omega^2 q^2 + \frac{ 1}{2} M \dot{Z}^2 - V(Z) + \lambda q(t) \Phi(t,Z(t))
\label{nonrelL1}
\end{eqnarray}
where we have dropped the term depending solely on the constant mass of the mirror $M$ and $V(Z)$ describes the potential energy of the mirror's motion.
The related Hamiltonian follows from a Legendre transform of (\ref{nonrelL1}) and is found to be
\begin{align}
	H = \frac{ 1}{2} \int dx \big( \Pi^2 + \Phi^{\prime 2} \big) + \frac{ p^2 }{ 2m} + \frac{1}{2} m \Omega^2 q^2 + \frac{ P^2}{2M} + V(Z) - \lambda q(t) \Phi(t, Z(t))
\label{nonrelH1}
\end{align}
The equations of motion are easily derived from (\ref{nonrelL1}) or (\ref{nonrelH1}) so we do not give them here.

\section{Multiple moving mirrors in the MOF model}
\label{sec:multiplemovingMOF}

In the previous sections we introduced a model for a  mirror whose scattering and reflective properties are described by an oscillator, the mirosc,  coupled bilinearly to the field.
In this section we extend the MOF model to include multiple spatially separated partially transmitting mirrors that interact mutually via the field. 

The Lagrangian for $N$ moving mirrors (possibly relativistically) with masses $M_a$ ($a=1,\ldots, N$) can be written as
\begin{align}
	L = \frac{1}{2}  \int dx \, \partial_\alpha \Phi \partial^\alpha \Phi + \sum_{a=1}^N \frac{ m_a}{2} \bigg( \frac{ \dot{q}_a^2 }{ \sqrt{ 1 - U_a^2 } }  - \Omega_a^2 q_a^2 \sqrt{ 1 - U_a^2 } \bigg) - \sum_{a=1}^N M_a \sqrt{ 1 - U_a^2 } + \sum_{a=1}^N \lambda_a \sqrt{1- U_a^2} \, q_a(t) \Phi(t, Z_a(t))
\end{align}
and the Euler-Lagrange equations of motion follow straightforwardly and are simply given by Eqs.~(\ref{eq:ELeom1})-(\ref{eq:ELeom3}) with all mirosc and worldline parameters and variables receiving a subscript $a$ to label the mirror. For completeness and for later use, the corresponding Hamiltonian is
\begin{align}
	H = \frac{1}{2} \int dx \big( \Pi^2 + (\partial_x \Phi)^2 \big) + \sqrt{ P^2 + M_{\rm eff}^2 (t) }
\end{align}
where the effective mass of the mirror has the same interpretation as before (i.e., mirror rest mass plus total internal energy) except now the total internal energy includes the energy of all $N$ mirosc's and their interaction energies with the field,
\begin{align}
	M_{\rm eff} (t) = M + \sum_{a=1}^N \bigg( \frac{ p_a^2 (t) }{ 2m_a} + \frac{1}{2} m_a \Omega_a^2 q_a^2(t) - \lambda_a q_a(t) \Phi (t, Z_a(t)) \bigg)  .
\end{align}
In the non-relativistic limit, the Lagrangian and the Hamiltonian are
\begin{align}
	L & = \frac{1}{2} \int dx \, \partial_\alpha \Phi \partial^\alpha \Phi + \sum_{a=1}^N \bigg( \frac{ 1}{2} m_a \dot{q}_a^2 - \frac{1}{2} m_a \Omega_a^2 q_a^2 + \frac{ 1}{2} M_a \dot{Z}_a^2 - V(Z_a) + \lambda_a q_a(t) \Phi (t, Z_a(t)) \bigg)
\label{eq:nonrelLforN} \\
	H & = \frac{1}{2} \int dx \big( \Pi^2 + (\partial_x \Phi)^2 \big) + \sum_{a=1}^N \bigg( \frac{ p_a^2 }{ 2 m_a } + \frac{1}{2} m_a \Omega_a^2 q_a^2 + \frac{ P_a^2 }{ 2 M _a} + V(Z_a) - \lambda_a q_a(t) \Phi(t, Z_a(t)) \bigg)
\label{eq:nonrelHforN}
\end{align}

In the remainder of this section, we investigate the scattering properties of incident radiation on two mirrors at rest. The equations of motion for the two-mirror MOF model are ($a=1,2$)
\begin{eqnarray}
	\partial_t^2 \Phi - \partial_x^2 \Phi &=& \sum_{a=1} ^2 \lambda_a q_a(t) \delta (x) \\
	m_a \ddot{q}_a + m_a \Omega_a^2 q_a &=& \lambda_a \Phi(t, 0)  .
\end{eqnarray}
Let a monochromatic plane wave of frequency $\omega$ be incident from the left so that
\begin{eqnarray}
	\Phi_{\omega L } (t,x) = e^{-i \omega t} \psi_{\omega L }(x)  .
\end{eqnarray}
The  part of the mode $\psi_{\omega L}(x)$ can be found using the linearity of the field equation from which the superposition principle allows us to write the contributions from multiple reflections and transmissions off of and through both mirrors as
\begin{eqnarray}
	\psi_{\omega L } (x) &=& \theta(-x) \left[ e^{i \omega x} + \left( R_1 + T_1 R_2 T_1 \sum_{n=0}^\infty (R_1 R_2)^n \right) e^{-i \omega x} \right] \nonumber \\
	&& + \theta(L - x) \theta(x) \left[ T_1 \sum_{n=0}^\infty (R_1 R_2)^n e^{i \omega x} + T_1 R_2 \sum_{n=0}^\infty (R_1 R_2)^n e^{-i \omega x} \right] \nonumber \\
	&&+ \theta(x-L) \left[ T_1 T_2 \sum_{n=0}^\infty (R_1 R_2)^n e^{i \omega x} \right]  .
\label{incmode2}
\end{eqnarray}
The geometric series can be summed for $| R_1 R_2 | < 1$ whereby
\begin{eqnarray}
	\sum_{n=0}^\infty (R_1 R_2 )^n = \frac{ 1}{ 1- R_1 R_2 }  .
\end{eqnarray}
To find the reflection and transmission coefficients in terms of the incident frequency $\omega$ we assume that the mirosc is in a steady-state evolution and  oscillates at the same frequency of the radiation so that
\begin{eqnarray}
	q_a(t) = A_a e^{-i \omega t}  .
\end{eqnarray}
The field is continuous at the locations of each mirror
\begin{eqnarray}
	\psi_{\omega L} (0^+) &=& \psi_{\omega L } (0^-)  \\
	\psi_{\omega L} (L^+) &=& \psi_{\omega L } (L^-)
\end{eqnarray}
and the discontinuity of the spatial derivative is to be consistent with the source of the field equation
\begin{eqnarray}
	- \psi^\prime _{\omega L } (0^+) + \psi^\prime _{\omega L} (0^- ) &=& \lambda_1 A_1 \\
	- \psi^\prime _{\omega L } (L^+) + \psi^\prime _{\omega L} (L^- ) &=& \lambda_2 A_2   ;
\end{eqnarray}
The mirosc amplitudes $A_1, A_2$ satisfy the mirosc equations of motion so that we have  six equations for the six unknowns $\{ R_a, T_a, A_a \}$ (note the subscript $a=1,2$). Thus, the reflection and transmission coefficients are
\begin{eqnarray}
	R_1 &=& \frac{ i \lambda_1^2 }{ 2 m_1 \omega (\Omega_1^2 - \omega^2 ) - i \lambda_1^2 } \\
	T_1 &=& 1 + R_1 \\
	R_2 &=& \frac{ i \lambda_2^2 \, e^{2 i \omega L} }{ 2m_2 \omega (\Omega_2^2 - \omega^2) - i \lambda_2 ^2 } \\
	T_2 &=& 1 + R_2 e^{-2 i \omega L}
\end{eqnarray}
and the amplitude of oscillation for the  miroscs are
\begin{eqnarray}
	A_1 &=& \frac{ \lambda_1 T_1 }{ m_1 (\Omega_1^2 - \omega^2) } \left( \frac{1+R_2}{ 1- R_1 R_2} \right) \\
	A_2 &=& \frac{ \lambda_2 T_2 }{ m_2 (\Omega_2 ^2 - \omega^2 ) } \left( \frac{ T_1 e^{i \omega L} }{ 1 - R_1 R_2 }  \right)  .
\end{eqnarray}
One can check that the identities $| R_a|^2 + |T_a|^2 = 1$ are indeed satisfied. The incident field mode (\ref{incmode2}) can then be written as
\begin{eqnarray}
	\psi_{\omega L } (x) &=& \theta( -x) \left[ e^{i \omega x} + \frac{ R_1 + R_2 + 2 R_1 R_2 }{ 1- R_1 R_2}  \,  e^{-i \omega x}  \right] \nonumber \\
	&& +  \theta(L-x) \theta(x) \left[ \frac{ T_1 }{ 1- R_1 R_2} \big( e^{i \omega x} + R_2 e^{- i \omega x} \big) \right] + \theta(x) \left[ \frac{ T_1 T_2 }{ 1 - R_1 R_2}  \, e^{i \omega x} \right]  .
\end{eqnarray}
When the mirror at $x=0$ is perfectly transmitting and the mirror at $x=L$ is perfectly reflecting the field mode is
\begin{eqnarray}
	\psi_{\omega L } (x) = \theta (L-x) \left( e^{i \omega x} - e^{i \omega (2L - x) } \right)  ,
\end{eqnarray}
which vanishes as $x \to L$, as expected. In the complementary case when the mirror at $x=0$ is perfectly reflecting  the field mode incident from the left is
\begin{eqnarray}
	\psi_{\omega L} (x) = \theta (-x) \left( e^{i \omega x} - e^{-i \omega x} \right)
\end{eqnarray}
as also expected. Hence, the MOF model describes the partially reflecting and transmitting properties of two, and generally more, mirrors.

\section{Classical mirror cooling with the MOF model}
\label{sec:cooling}

In this Section, we show how the MOF model can be used to describe mirror cooling within a completely classical context. In a following paper, we discuss quantum effects in mirror cooling using the MOF model \cite{BGH_mircool}.

The setup is as follows. Consider a cavity formed by two mirrors. We take one of the mirrors to be fixed at the origin and perfectly reflecting so that the (classical scalar) field satisfies Dirichlet boundary conditions, $\Phi (t, 0)=0$. As this fixed and perfectly reflecting mirror will, by assumption, possess no dynamics then we will model the second mirror by the MOF model. This second mirror possesses a mirosc internal degree of freedom and will be free to move in response to the forces imparted by the field. The motion of this second mirror is assumed to be small relative to the size of the cavity, $L$, and to move on a time-scale much longer than all other time scales in the problem. The partial reflectivity of the second mirror allows, for example, a laser field, generated by an external source $J_{\rm ext}(x^\alpha)$, to couple to the cavity.

\subsection{Arbitrary bilinear coupling strength}

The MOF Lagrangian for the system described in the previous paragraph is given by Eq.~(\ref{lagrangian00})
\begin{align}
\label{}
	L & =  \frac{1}{2} \int dx \big( \dot{\Phi}^2(x^\alpha) - \Phi^{\prime 2}(x^\alpha) + 2 J_{\rm ext}(x^\alpha)  \Phi (x^\alpha) \big) +  \frac{m}{2} \big( \dot{q}^2(t) - \Omega^2 q^2(t) \big) +  \frac{M}{2} \big( \dot{Z}^2(t) - \Omega_0^2 Z^2(t) \big) + \lambda q(t)  \Phi(t, L +Z(t))
\end{align}
where we have included an external source $J_{\rm ext}(x^\alpha)$ for the field and the second mirror (the dynamical one) has coordinates $x = L + Z(t)$ and moves within a harmonic potential with natural frequency $\Omega_0$.
The Euler-Lagrange equations for the field, the mirosc, and the coordinates of the movable mirror are
\begin{align}
\label{}
	\partial_t^2 \Phi (x^\alpha)  - \partial_x^2 \Phi (x^\alpha)  & = J_{\rm ext} (x^\alpha) + \lambda q(t)  \delta(x- L - Z(t))
\label{eq:eomPhi10} \\
	\ddot{q}(t) + \Omega^2 q(t) & = \frac{\lambda}{m} \Phi(t, L + Z(t))
\label{eq:eomMirosc10}  \\
	\ddot{Z}(t) + \Omega_0^2 Z(t)  & = \frac{\lambda}{M} q(t) \, \partial_x \Phi (t, L+ Z(t))  .
\label{eq:eomZ10}
\end{align}

Our first step will be to solve (\ref{eq:eomPhi10}) for the field and eliminate its appearance in the remaining equations of motion. Assuming that there is no initial field present \cite{footnote2}, so that $\Phi$ is generated by $J_{\rm ext}$ and by interactions with the remaining degrees of freedom, then the solution to (\ref{eq:eomPhi10}) is
\begin{equation}
\label{eq:Phi10}
\Phi(x^\alpha) =  \int d^2 x' \, G(x^\alpha; x'^\alpha) J_{\rm ext} (x'^\alpha) + \lambda \int dt' \, G(x^\alpha ; t', L + Z(t')) q(t')
\end{equation}
where the retarded Green's function $G(x^\alpha; x'^\alpha)$ for the field subject to Dirichlet boundary conditions at the fixed mirror is given by
\begin{align}
	G (t, x; t', x') = \frac{1}{2} \theta(t-t') \bigg[ \theta \bigg( \frac{1}{2} (t-t')^2 - \frac{1}{2} (x-x')^2 \bigg) - \theta \bigg( \frac{1}{2} (t-t')^2 - \frac{1}{2} (x+x')^2 \bigg) \bigg] .
\label{eq:Green2}
\end{align}
Note that if $x=x'=L>0$ then
\begin{align}
	G(t, L; t', L) = {} & \frac{1}{2} \big[ \theta( t- t') - \theta ( t-t' - 2L) \big]   .
\label{eq:Green3}
\end{align}	

Substituting (\ref{eq:Phi10}) into the remaining equations (\ref{eq:eomMirosc10}) and (\ref{eq:eomZ10}) gives
\begin{align}
	\ddot{q}(t) + \Omega^2 q(t) = {} & \frac{ \lambda}{m} F_{\rm ext} (t, L+Z(t))  + \frac{\lambda^2}{m} \int dt' \, G(t, L + Z(t); t', L + Z(t')) q(t')
\label{eq:eomMirosc11} \\
	\ddot{Z}(t) + \Omega_0^2 Z(t) = {} & \frac{ \lambda }{ M } q(t) \partial_x F_{\rm ext} (t, L+Z(t)) + \frac{ \lambda^2 }{ M } q(t) \int dt' \, \partial_x G (t, L+ Z(t) ; t', L + Z(t')) q(t')
\label{eq:eomZ11}
\end{align}
where
\begin{align}
	F_{\rm ext} (x^\alpha) \equiv \int d^2 x' \, G(x^\alpha; x'^\alpha) J_{\rm ext} (x'^\alpha)
\end{align}
is the propagated external source for the field.

Next, we solve for the mirosc variable, $q(t)$. At this point we can take advantage of the assumption that $Z(t) \ll L$ so that the typical amplitude of the mirror's motion is much smaller than the size of the cavity. This implies we can write the solution for the oscillator perturbatively as $q = q_0 + q_1 + \cdots$ where $q_n = O(Z^n)$. The equation of motion for the leading order mirosc dynamics is
\begin{align}
\label{eq:eomMirosc12}
	\ddot{q}_0 (t) + \Omega^2 q_0(t) - \frac{ \lambda^2}{m} \int dt' G(t, L ; t', L ) q_0(t') = \frac{\lambda }{m} F_{\rm ext}(t, L)   .
\end{align}
 The solution to (\ref{eq:eomMirosc12}) is given by (again, ignoring homogeneous solutions)
\begin{align}
	q_0 (t) =  \lambda \int_{-\infty}^\infty dt' \, D(t-t') F_{\rm ext} ( t', L)
\label{eq:eomMirosc13}
\end{align}
where the kernel $D(\tau)$ is found to be
\begin{align}
	D(\tau) = - \int _{-\infty}^\infty \frac{ d\omega }{ \pi } \, \frac{ \omega e^{- i \omega \tau} }{ 2 m \omega (\omega^2 - \Omega^2 ) + i \lambda^2 ( 1 - e^{2 i \omega L } ) } \equiv \int_{-\infty}^{\infty} \frac{d \omega}{2\pi} \ e^{- i \omega \tau} \tilde{D}(\omega) .
\end{align}
The equation of motion for the first order perturbative correction to the mirosc dynamics is
\begin{align}
	\ddot{q}_1 (t) + \Omega^2 q_1 (t) - \frac{ \lambda^2 }{ m } \int dt' \, G(t, L; t', L) q_1 (t') = {} & \frac{\lambda }{m} \partial_x F_{\rm ext}(t, L) Z(t) + \frac{\lambda^2}{m} \int dt' \big[ Z(t) \partial_{x} + Z(t') \partial_{x'} \big] G(t, L; t', L) q_0(t')  .
\label{eq:eomMirosc14}
\end{align}
The right side of (\ref{eq:eomMirosc14}) simplifies somewhat since (\ref{eq:Green2}) implies that
\begin{align}
	\partial_x G(t, L; t', L) & = \frac{1}{2} \delta (t-t' - 2L) = \partial_{x'}  G(t, L ; t', L)
\end{align}
and so (\ref{eq:eomMirosc14}) can be written as
\begin{align}
	\ddot{q}_1 (t) + \Omega^2 q_1 (t) - \frac{ \lambda^2 }{ m } \int dt' \, G(t, L; t', L) q_1 (t') = {} & \frac{\lambda }{m} \partial_x F_{\rm ext} (t, L) Z(t) + \frac{\lambda^2}{2m} \big( Z(t) + Z(t-2L) \big) q_0 (t-2L) .
\label{eq:eomMirosc15}
\end{align}
Thus, the solution to (\ref{eq:eomMirosc15}) is given by
\begin{align}
	q_1(t) = \int_{-\infty}^\infty dt' \, D(t-t') \bigg[ \lambda \partial_x F_{\rm ext} (t', L) Z(t') + \frac{ \lambda^2}{2} \big( Z(t') + Z(t'-2L) \big) q_0 (t'-2L) \bigg] .
\label{eq:eomMirosc16}
\end{align}

Next, we expand the equation of motion for the worldline to leading order in $Z(t)$ to find
\begin{align}
	M \ddot{Z}(t) + M \Omega_0^2 Z(t)  = {\cal F} [ Z(t) ]
\label{eq:eomMirror10}
\end{align}
where ${\cal F}[Z(t)]$ accounts for the external forces and backreaction from the cavity field and mirosc, and is given by
\begin{align}
\label{eq:effsourceF1}
  {\cal F} [ Z(t) ] = & \lambda q_0(t) \bigg( \partial_x F_{\rm ext} (t, L) + \frac{\lambda}{2} q_0( t- 2 L) \bigg) +  \lambda q_0(t) \bigg( \partial^2_x F_{\rm ext} (t,L) - \frac{\lambda}{2} \dot{q}_0( t- 2 L) \bigg) Z(t) \nonumber
  \\
 + &  \lambda q_0(t) \bigg( - \frac{\lambda}{2} \dot{q}_0 (t-2L) Z(t)  - \frac{\lambda}{2} q_0( t- 2 L) \dot{Z}(t- 2 L) \nonumber
 - \frac{\lambda}{2} \dot{q}_0( t- 2 L){Z}(t- 2 L)  \bigg)
 \\
 + & \lambda q_1(t) \bigg( \partial_x F_{\rm ext} (t,L) + \frac{\lambda}{2} q_0( t- 2 L) \bigg) +  \frac{\lambda^2}{2} q_0(t)   q_1(t - 2L ) \nonumber.
\end{align}
We find that the general motion of the mirror as influenced by the cavity field is described by
a delay integro-differential equation.

The backreaction terms above will be shown to lead to several effects.
First, the driven field will build up in amplitude inside the cavity formed
by the perfect mirror and the mirror-oscillator. This will lead to a spatially
varying radiation pressure and a shift in the frequency of the mirror's
mechanical motion. Next, depending on the equilibrium position of the mirror
the cavity field can either accept from or donate energy to the mirror's motion
arising from retardation effects (see e.g. \cite{MarGir} for a detailed explanation of
cooling due to retardation).
Finally, non-Markovian effects will be present which show how the mirror's
motion in the past influences its future movements, these effects are accounted
for in time-delayed and integral terms.

\subsection{The weak-coupling limit}

As an example application of these equations we will explore mirror cooling in the weak coupling limit i.e. $\lambda^2 / (m \Omega^3) \ll 1$.
For many systems of physical interest there exists a large separation between the values of the cavity
frequency and the oscillation frequency $\Omega_0$, which allows for a multiple time-scale analysis. In the following we will assume that the
cavity frequency, the mirosc's frequency $\Omega$, and the pump frequency $\Omega_D$ are all much larger than the frequency of the mirror's
mechanical motion $\Omega_0$.  Under these circumstances we may time-average the mirror's equation of motion in (\ref{mirror-equation}) over the pump period $2 \pi / \Omega_D$. Since the mirror's
mechanical motion is very slow compared to this pumping time-scale its trajectory can be safely factored out of any time-averaging integrals so that
\begin{align}
	\big\langle \big\langle Z(t) (\cdots) \big\rangle \big\rangle & \equiv \frac{1}{T} \int_t^{T+t} \!\!\! dt' \, Z(t') (\cdots )  \approx \frac{ Z(t) }{ T } \int_t^{T+t} \!\!\! dt' (\cdots )
\end{align}
where $\langle\langle \cdot \rangle \rangle$ denotes the time average.

Let the external source of the field be given by
\begin{align}
	J_{\rm ext} ( x^\alpha) = A \cos \Omega_D t
\end{align}
so that
\begin{align}
	F_{\rm ext} (x^\alpha) = A \int_{t_i}^\infty dt' \int_0^\infty dx' \, G (t, x ; t', x') \cos \Omega_D t'
\end{align}
where we take the initial time to be at $t= t_i$ and at the end of the calculation take the limit $t_i \to -\infty$. Performing the spacetime integration gives
\begin{align}
	F_{\rm ext} (t, L) & = \alpha e^{- i \Omega_D t}  + {\rm c.c.}
\end{align}
where
\begin{align}
	\alpha & = \frac{ A }{ 2 \Omega_D^2 } \big( e^{i \Omega_D L } - 1 \big)
\end{align}
The first two spatial derivatives of $F_{\rm ext}
$ evaluated at $x=L$ are similarly evaluated giving
\begin{align}
	\partial_x F_{\rm ext} (t, L) & = \alpha' e^{-i \Omega_D t} + {\rm c.c.} \\
	\partial_x^2 F_{\rm ext} (t, L) & = i \Omega_D \alpha' e^{-i \Omega_D t} + {\rm c.c.}
\end{align}
where
\begin{align}
		\alpha' & = \frac{ i A}{ 2 \Omega_D } e^{ i \Omega_D L}  .
\end{align}
In addition, we can also derive the explicit form for $q_0(t)$ given the expression for the external source

\begin{equation}
\label{ }
q_0(t) = \lambda \alpha \tilde{D}(\Omega_D) e^{- i \Omega_D t} + c.c..
\end{equation}

Using these expressions we shall evaluate the time-average of (\ref{eq:eomMirror10}) over the pump period, $T = 2 \pi / \Omega_D$. The time-average of the terms independent of $q_1(t)$ in (\ref{eq:effsourceF1}) are easily evaluated.
However, the time-average of the terms in (\ref{eq:eomMirror10}) containing $q_1(t)$ requires some elaboration. First, we write $q_1(t)$ in (\ref{eq:eomMirosc16}) with $Z(t)$ and $D(t-t')$ replaced by their Fourier transforms,
\begin{equation}
\label{ }
q_1(t) =  \lambda \int_{-\infty}^\infty  \frac{ d \nu}{2 \pi}  \frac{ d \omega}{ 2 \pi} \int dt' \,  Z (\nu) \tilde{D}(\omega) e^{ - i \nu t' - i \omega(t-t') } \bigg[ \partial_x F_{\rm ext}(t',L) + \frac{\lambda}{2} \big( 1 + e^{ i 2 \nu L }  \big) q_0(t'- 2 L) \bigg].
\end{equation}
Evaluating the $t'$ integral and then integrating over $\omega$ gives
\begin{align}
\label{q1-FT}
q_1(t) = {} &  \lambda \int_{-\infty}^\infty \frac{ d \nu}{2 \pi}   Z (\nu)  e^{ - i \nu t} \bigg\{ e^{ -i \Omega_D t}  \tilde{D}(\nu + \Omega_D)  \bigg[ \alpha'  + \frac{\lambda^2}{2} \big( 1 + e^{ i 2 \nu L }  \big)  \alpha \tilde{D}(\Omega_D) e^{ i 2 \Omega_D L} \bigg] \nonumber \\
& {\hskip1.2in}+ e^{ i \Omega_D t}  \tilde{D} ( \nu -\Omega_D )  \bigg[ \alpha'{}^*  + \frac{\lambda^2}{2} \big( 1 + e^{ i 2 \nu L }  \big)  \alpha^* \tilde{D}^*(\Omega_D) e^{ -i 2 \Omega_D L} \bigg]
\bigg\}.
\end{align}
In the weak coupling limit, $\lambda^2 \ll m \Omega^3$, the effect of the cavity field on the (forced) mirror motion is sufficiently small that the mirror will continue to oscillate at a frequency nearly equal to $\Omega_0$. Consequently, we expect $Z(\nu)$ to be sharply peaked for frequencies $\nu \sim \Omega_0$. Since $L$ is inversely proportional to the cavity period and $\nu \sim \Omega_0$ then it follows that $\nu L \ll 1$ and $\Omega_D \gg \nu$. Therefore, expanding all terms but $Z(\nu)$ in the integrand of (\ref{q1-FT}) for $\nu$ near zero \footnote{We do not expand $Z(\nu)$ itself since we wish to find the equation of motion for the time-averaged $Z(t)$ that is self-consistent with the evolution of the field and mirosc.}
\begin{align}
\label{q1-FT2}
q_1 (t) \approx  \lambda \bigg\{ e^{ -i \Omega_D t}  D(\Omega_D)  \bigg[ \alpha'  + \lambda^2    \alpha \tilde{D}(\Omega_D)  e^{ i 2 \Omega_D L} \bigg]+
e^{ i \Omega_D t}  D(-\Omega_D )  \bigg[ \alpha'{}^*  + \lambda^2   \alpha^* \tilde{D}^*(\Omega_D)  e^{ -i 2 \Omega_D L} \bigg]
\bigg\} Z(t).
\end{align}
Lastly, since the mirror moves on a time scale ($\sim 1/\Omega_0$) much longer than the round trip travel time of light in the cavity ($\sim 2L$) then $\Omega_0 L \ll 1$ and we may expand all delay terms about their instantaneous values so that, for example, $Z(t-2L) = Z(t) -2L \dot{Z}(t) + O( ( \Omega_0 L)^2)$. Using (\ref{q1-FT2}) one may then easily compute the time-average of the terms depending on $q_1(t)$ in (\ref{eq:effsourceF1}). Putting everything together, and remembering to expand the delay terms as discussed above, we find that (\ref{eq:eomMirror10}) becomes
\begin{equation}
\label{mirror-equation}
	M \ddot{Z}(t) + \Gamma(L)\dot{Z}(t)+ M (\Omega^2_0- \Delta \Omega^2(L) ) Z(t) = F_{\rm rad}(L) ~,
\end{equation}
which is simply the equation for a forced, damped harmonic oscillator with mass $M$, frequency $[\Omega_0^2 - \Delta \Omega^2 (L)]^{1/2}$, and damping coefficient $\Gamma(L)$. Notice that the latter two quantities depend explicitly on the length of the cavity.
After time-averaging the explicit form for the radiation pressure is given by

\begin{align}
\label{}
 F_{\rm rad} (L) = \lambda^2 \alpha \tilde{D}( \Omega_D) \bigg(\alpha^{'*} + \frac{\lambda^2}{2} \alpha^* \tilde{D}^* (\Omega_D) e^{-i2\Omega_D L} \bigg),
\end{align}
the frequency shift is given by

\begin{align}
\label{}
M \Delta \Omega^2(L) = & - i \Omega_D \lambda^2 \alpha \tilde{D}(\Omega_D)\bigg(\alpha^{'*} + \frac{3\lambda^2}{2} \alpha^* \tilde{D}^* (\Omega_D) e^{-i2\Omega_D L} \bigg)
\nonumber \\
& + \lambda^2 \tilde{D}(\Omega_D) \bigg( \alpha' + \lambda^2 \alpha \tilde{D}(\Omega_D) e^{ i 2 \Omega_D L} \bigg)
\bigg(\alpha^{'*} + \lambda^2 \alpha^* \tilde{D}^* (\Omega_D) \cos 2\Omega_D L \bigg),
\end{align}
and the damping coefficient is given by

\begin{align}
\label{}
\Gamma(L) = \frac{\lambda^4}{2} |\alpha \tilde{D}(\Omega_D) |^2 \cos 2 \Omega_D L.
\end{align}

\begin{figure}
\includegraphics[width=0.9\columnwidth]{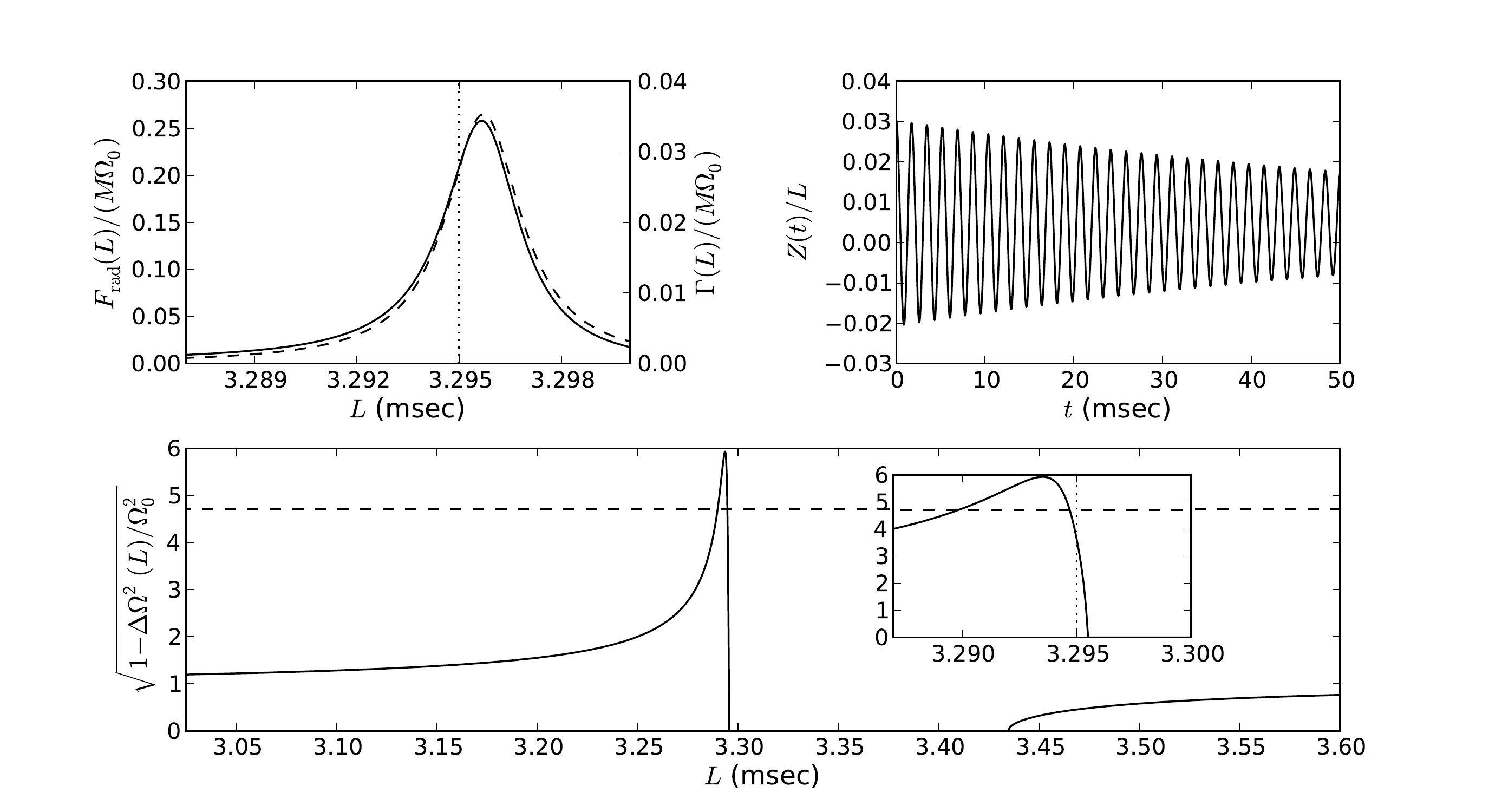}
\caption{Plots of the force on the mirror from radiation pressure (top left, solid line), the damping constant $\Gamma(L)$ (top left, dashed line), and the fractional change in the mirror's natural frequency (bottom) as a function of the unperturbed cavity length $L$. The dotted vertical line indicates the value of $L=3.295\,$ms. The dashed line in the bottom panel shows an estimate of where the weak coupling approximation begins to break down. The inset shows the fractional change in frequency for $L \in [ 3.287, 3.300]\,$ms. The top right panel shows the evolution of the perturbed trajectory as a function of time with $L=3.295$ms. The parameter values, in units where $c=1$, used to make these plots are as follows: $(m,M) = (1, 2\,000)\,$kg, $(\Omega_0, \Omega_D, \Omega) = (1, 300 \pi \approx 942.5, 942)\,$Hz, $\lambda=5\,000\,$s${}^2$, and $A=10\,$s${}^{-2}$.}
\label{fig:cooling}
\end{figure}

Fig.\,\ref{fig:cooling} shows the force on the mirror due to the resulting radiation pressure $F_{\rm rad}(L)$ (solid line) and the damping constant $\Gamma(L)$ (dashed line) as a function of the movable mirror's unperturbed position $L$ from the static mirror at the origin. The parameter values chosen for these plots are given in the corresponding figure caption. We observe that when the cavity is pumped by an external source, the field energy inside builds up and results in a  force from radiation pressure $F_{\rm rad}(L)$ that varies depending on the size of the unperturbed cavity.

The gradient of the radiation pressure and the Markov approximation of the integral terms, i.e. those terms containing $q_1(t)$,
leads to a shift in the oscillation frequency of the
mirror's center of mass motion. These optical spring effects are quantified by the term $\Delta \Omega(L)$ and changes depending on $L$. The bottom panel of Fig.\,\ref{fig:cooling} shows the fractional change in the mirror's natural frequency $[1-\Delta\Omega^2(L)/\Omega^2_0]^{1/2}$, which can become imaginary precisely where the real part goes to zero in that plot.
An important point to note here is that our weak coupling approximation is valid when $| \Omega_0 - \Delta \Omega(L) | \ll \Omega_D$. For the values indicated in the figure caption and with $L=3.295$ms we see that the mirror's modified natural frequency satisfies $\sqrt{\Omega_0^2 - \Delta \Omega^2 } \approx 3.64 {\rm Hz} << \Omega_D$, which is consistent with the weak coupling approximation. It is also important to mention that the mirror's motion can become unstable when the mirror is to the right of the resonance, namely, the mirror's spring constant, i.e. $K = M(\Omega^2-\Delta \Omega^2(L))$, becomes negative as shown in Fig.\,\ref{fig:cooling}.
For the parameter values given in the caption of Fig.\,\ref{fig:cooling}, the moving mirror's motion is damped, the top right panel in Fig.\,\ref{fig:cooling}, and exemplifies the ``cooling'' aspect of this classical system to dissipate its input energy into the cavity field.

\section{Reduction of MOF model to Models with NX-coupling}
\label{sec:nxcoupling}

In the previous section we used the MOF model in the classical regime to describe the damped motion of one mirror of a cavity forced by interactions with an external (laser) field. In the corresponding quantum theory of mirror cooling, one usually models the interaction between the mirror and the field by the radiation pressure $\sim \hat{N} \hat{x}$ where $\hat{N}$ is the number operator of quanta (photons) impinging on the mirror's surface and $\hat{x}$ is the position operator of the mirror \cite{MarGir,Caves80,Caves81}.
We will refer to this interaction as ``$Nx$-coupling.''

The basic motivation for this type of interaction can be easily understood by considering the Hamiltonian for
a single cavity mode of the form, $H_{\rm cav} \sim \omega_{\rm cav}(L) a^\dag a$, where $a$ ($a^\dag$) is the annihilation
(creation) operator for field quanta and $\omega_{\rm cav}(L)$ is the frequency of a cavity mode of size $L$. Since the frequency of the cavity modes scales as the inverse cavity length $\omega_{\rm cav}(L) \sim 1/L$,
when we allow the cavity length to vary by a small amount $x$ the frequency is perturbed to leading order as $\omega_{\rm cav}(L+x) \approx  \omega_{\rm cav}(L) (1 - x/L+ ...)$. For small cavity length changes the Hamiltonian becomes $H_{\rm cav} \approx \omega_{\rm cav}(L)(1 - x/L) a^\dag a$

In this section, we show how the quantum MOF model relates to models with $Nx$-coupling. In doing so, we highlight the assumptions that must be made to connect the two models. We thereby demonstrate that the MOF model should be an improvement of the oft used background field approximation for the cavity field \cite{Kimble,BraKha}
In particular, the MOF model should be very useful for studying optomechanical systems having low numbers of cavity photons where quantum effects can become quite interesting and important. 

Consider a cavity formed by two mirrors. As in Section \ref{sec:cooling}, we take the first mirror (at $x=0$) to be fixed for all time and perfectly reflecting so that the field satisfies Dirichlet boundary conditions at the origin. We assume the second mirror to be partially transmitting and dynamical with small perturbations to its equilibrium position at $x=L>0$. The second mirror will be described by the MOF model.
Recall the Hamiltonian (\ref{nonrelH1}) for a slowly moving mirror
\begin{align}
\label{Free-Hamiltonian}
	H =  \frac{1}{2} \int dx \, \big( \Pi^2(x^\alpha) + (\partial_x \Phi (x^\alpha))^2 - 2 J_{\rm ext}(x^\alpha) \Phi(x^\alpha) \big) + \frac{p^2 (t)}{2 m} + \frac{1}{2} m \Omega^2 q^2(t)  +  \frac{P^2(t)}{2 M} + V(Z(t))  - \lambda q(t) \Phi (t, L + Z(t))
\end{align}
where we have included an external source $J_{\rm ext} (x^\alpha)$ for the field. We shall show, by making a number of assumptions, that the interaction component of the above Hamiltonian
\begin{align}
\label{interaction}
	H_{\rm int} = - \lambda \int dx \ q(t) \Phi(t, L + Z(t) )
\end{align}
can be reduced to the $Nx$-coupling.

The internal physics of the mirror for many standard radiation pressure cooling calculations is accounted for phenomenologically through the introduction of a cavity quality factor which accounts for the dissipation of field energy from within the cavity.
In distinction, the MOF model accounts for the detailed information of the mirror's internal dynamics. We first will solve for the mirosc to find its effect on the mirror's motion. In this way we trade the microscopic information about the mirror for a macroscopic description in terms of the mirror's susceptibility, which will establish the link between $Nx$-coupling and the MOF model.

The Heisenberg equations of motion for the field (\ref{eq:ELeom1}) and the mirosc (\ref{eq:ELeom2}) variables are
\begin{align}
\label{HEOMF}
	\partial_t^2 \Phi(x^\alpha) - \partial_x^2 \Phi(x^\alpha) & = \lambda q(t) \delta(x -L - Z(t))  + J_{\rm ext} (x^\alpha) \\
\label{HEOMO}
	\ddot{q}(t) + \Omega^2 q(t) & = \frac{\lambda}{m} \, \Phi(t, L +Z(t)).
\end{align}
We can eliminate the field's explicit dependence on the mirosc $q(t)$ by solving (\ref{HEOMO}) and plugging the solution into the wave equation (\ref{HEOMF}). In the regime where the mirosc evolves adiabatically so that $| \ddot{q}| \ll |\Omega^2 q|$  the approximate solution to (\ref{HEOMO}) is given by
\begin{equation}
\label{solution}
	q(t)  \approx \frac{ \lambda }{ \kappa } \Phi (t, L + Z(t)) - \frac{ \lambda }{ \kappa } \frac{ \ddot{\Phi}(t, L + Z(t)) }{ \Omega^2 } + \cdots
\end{equation}
where the second term on the right side is a correction to the leading order, instantaneous solution and is due to the fact that the full mirosc dynamics is generally non-Markovian. This can be seen from the general solution of (\ref{HEOMO})
\begin{align}
	q(t) = q_h(t) + \frac{\lambda}{m} \int dt' \, g_{\rm ret} (t; t') \Phi (t', L + Z(t'))
\label{Q_solution}
\end{align}
where $q_h(t)$ is the homogeneous solution and $g_{\rm ret} (t;t')$ is the retarded Green's function for the mirosc
\begin{align}
	g_{\rm ret} (t; t') = \theta (t-t') \frac{ \sin \Omega (t-t') }{ \Omega }  .
\end{align}
More specifically, the mirosc receives contributions from the past as implied by the integral in (\ref{Q_solution}). However, the approximation (\ref{solution}) is valid if the mirosc degree of freedom is ``light'' thus responding {\it nearly} instantaneously to external influences.

Substituting the approximate mirosc solution (\ref{solution}) into the wave equation (\ref{HEOMF}) gives the effective dynamics for the field
\begin{align}
\label{eq:effPhi1}
	\partial_t^2 \Phi(x^\alpha) - \partial_x^2 \Phi (x^\alpha) = J_{\rm ext} (x^\alpha) + \frac{ \lambda^2 }{ \kappa} \Phi (t, L + Z(t)) \delta (x - L -Z(t)) - \frac{ \lambda^2 }{ \kappa \Omega^2 } \ddot{\Phi} (t, L + Z(t)) \delta (x - L - Z(t)) + \cdots .
\end{align}
Notice that (\ref{eq:effPhi1}) can be derived from the following effective Lagrangian
\begin{align}
	L_{\rm eff}   = \frac{1}{2} \int dx \big( \partial_\alpha \Phi \partial^\alpha \Phi + 2 J_{\rm ext} \Phi \big) + \frac{ \lambda^2}{ 2\kappa} \int dt \, \Phi^2 (t, L + Z(t)) + \frac{ \lambda^2 }{ 2 \kappa \Omega^2 } \int dt \, \dot{\Phi}^2 (t, L + Z(t)) + \cdots + \frac{ 1}{2} M \dot{Z}^2 - V(Z)
\end{align}
where $\cdots$ denotes the higher order terms in (\ref{solution}). The interaction Hamiltonian corresponding to the above effective Lagrangian is found to be
\begin{align}
	H_{\rm eff \, int} = - \frac{ \lambda^2 }{ 2 \kappa } \Phi^2 (t, L + Z(t)) + \frac{ \lambda^2 }{ 2 \kappa \Omega^2 } \Pi^2 (t, L+ Z(t)) + \cdots .
\label{eq:effintH1}
\end{align}

Assuming that $Z(t) \ll L$, we may expand the effective interaction Hamiltonian in (\ref{eq:effintH1}) to find
\begin{align}
\label{ }
	H_{\rm eff \, int} = H^{(0)}_{\rm eff \, int} + H^{(1)}_{\rm eff \, int} + O(Z^2)
\end{align}
where
\begin{align}
	H^{(0)}_{\rm eff \, int} & = - \frac{ \lambda^2}{ 2\kappa} \Phi ^2 (t, L) + \frac{ \lambda^2 }{ 2 \kappa \Omega^2 } \Pi^2 (t, L)  + \cdots \\
	H^{(1)} _{\rm eff \, int} & = - \frac{ \lambda^2}{ \kappa} Z(t) \Phi(t, L) \partial_x \Phi (t, L) + \frac{ \lambda^2 }{ \kappa \Omega^2 } Z(t) \Pi (t, L) \partial_x \Pi(t, L) + \cdots  .
\label{eq:H1effint1}
\end{align}
Notice that the leading order interaction Hamiltonian is independent of $Z(t)$ so that it exerts no force on the movable mirror. In fact, one can group $H^{(0)}_{\rm eff \, int}$ with the free Hamiltonian for the field  that, when taken together, describes the free evolution of the field in a cavity where one mirror is fixed at $x=0$ and perfectly reflecting and the other mirror is fixed at $x=L$ but partially transmitting. The remaining terms in the effective interaction Hamiltonian   describe the perturbative response of the second mirror to its coupling with the field and vice versa.

To leading order in $Z(t)$, we can express the field in terms of a homogeneous solution via the cavity's normal modes and in terms of the external source $J_{\rm ext}$,
\begin{equation}
\label{ }
\Phi(x^\alpha) \approx \sum_k N_k \big( a_k u_k(x) e^{- i \omega_k t} + {\rm H.c.} \big) + \int d^2 x' \, G^{\rm cav}_{\rm ret} (x^\alpha, x'^\alpha) J_{\rm ext} (x'^\alpha)
\end{equation}
where ${\rm H.c.}$ is the Hermitian conjugate of the preceding terms, $u_k(x)$ are the normal modes of the cavity and satisfy
\begin{equation}
\label{ }
\bigg( \partial_x^2 + \omega_k^2  + \frac{\lambda^2}{\kappa} \delta(x- L) \bigg) u_k(x) = 0
\end{equation}
such that $u_k (0) = 0$ since the mirror at $x=0$ is perfectly reflecting. The retarded Green's function here satisfies

\begin{align}
\label{}
 \bigg( \partial_x^2 + \omega_k^2  + \frac{\lambda^2}{\kappa} \delta(x- L) \bigg) G^{\rm cav}_{\rm ret}(\omega_k; x,x') = -\delta(x-x')
 \end{align}
 with Dirichlet boundary conditions at the origin $G^{\rm cav}_{\rm ret}(\omega_k; x,0) = 0$ and $G^{\rm cav}_{\rm ret}(\omega_k; 0,x') = 0$.

Also, $N_k$ is chosen so that $[ \Phi(t,x), \Pi(t,x')] = i \hbar \delta(x-x')$ for $x$ and $x'$ greater than zero. These commutation relations require $a_k$ and $a^\dag_k$ to be annihilation and creation operators, respectively.
For the following we focus entirely on the component of the interaction Hamiltonian coming from the field {\rm inside} the cavity. The field outside of the cavity gives rise to a constant and position independent radiation pressure that only yields a shift in the equilibrium position of the mirror at $x=L$.

If the cavity is pumped by a laser beam with a frequency slightly detuned from one of the cavity resonances
and if the cavity quality factor is large then the cavity field, represented as a mode sum, can be approximated well by a single mode. Expressing the field in terms of the fundamental cavity resonance we find, at linear order $Z(t)$, that the interaction Hamiltonian (\ref{eq:H1effint1}) is given by
\begin{equation}
\label{ }
 H^{(1)}_{\rm int} \approx -  \frac{\lambda^2}{\kappa} Z(t)
\bigg[ N_k \big( a_k u_k(L) e^{- i \omega_k t} + {\rm H.c.} \big) + \tilde{F}_{\rm ext} (t, L) \bigg] \bigg[ N_k \big( a_k u'_k(L) e^{- i \omega_k t} + {\rm H.c.} \big) + \partial_x  \tilde{F}_{\rm ext} (t, L) \bigg] + \cdots
\end{equation}
where $\cdots$ refers to corrections arising from time derivatives of the field appearing in (\ref{eq:H1effint1}) and
\begin{align}
	\tilde{F}_{\rm ext} (x^\alpha) \equiv \int d^2 x' \, G^{\rm cav}_{\rm ret} (x^\alpha; x'^\alpha) J_{\rm ext} (x'^\alpha)  .
\end{align}
For many systems of interest the frequency of the fundamental cavity mode is much larger than the typical frequency of the mirror's motion (i.e. $\Omega_0/\omega_k \ll1$). Under such conditions the mirror's position changes adiabatically over many oscillations of the cavity field allowing a time average (denoted by double angled brackets) of the effective interaction Hamiltonian
\begin{equation}
\label{ }
	\langle\!\langle H^{(1)}_{\rm int}  \rangle\!\rangle = \frac{1}{NT}\int_0^{NT} \!\!\! dt \, H^{\rm (1)}_{\rm int}     .
\end{equation}
Here, $T$ is the period of the cavity's fundamental mode and $N$ is a large integer such that $(2\pi)/\Omega_0 \gg N T$. Since $Z(t)$ is approximately constant over the entire integration range it can be taken outside of the time-average giving
\begin{equation}
\label{ }
	\langle\!\langle H^{(1)}_{\rm int} \rangle\!\rangle  \approx - \frac{ \lambda^2 }{ \kappa} Z(t) \tilde{F}_{\rm ext} (t, L) \partial_x \tilde{F}_{\rm ext} (t, L) - \bigg(  \frac{\lambda^2}{2 \kappa}  |N_k|^2  u'_k(L) u^*_k(L) \bigg) Z(t) a^\dag_k a_k  + {\rm H.c.}
 \end{equation}
This step is equivalent to taking the rotating wave approximation. The key point is that the first term on the right side is a classical radiation pressure originating solely from the external source while the second term is a quantum mechanical radiation pressure and is, in fact, the $Nx$-coupling. 

Before concluding this section, we collect the main assumptions  used in relating the MOF model to the phenomenological radiation pressure interaction Hamiltonian. The assumptions are as follows:
 \begin{itemize}
	\item The movable mirror is only ever slightly perturbed from its otherwise equilibrium position at $x=L$;
	\item The cavity frequency is much less than the natural frequency of the mirosc;
	\item The cavity has a high quality factor;
	\item The cavity is pumped by a laser at a frequency slightly detuned from one of the cavity resonances; and
	\item The cavity frequency is much greater than the typical timescale associated with the mirror's motion (i.e., the natural period if in a harmonic trap)
\end{itemize}
Under these assumptions we have shown that the effective interaction between the mirror and the cavity field is given by an $Nx$-coupling. It is possible that the $Nx$-coupling can be obtained using a different setup and assumptions. However, our purpose here is not to elucidate all the ways that the $Nx$-coupling can be derived from the MOF model but rather to show that it \textit{can} be derived  from a microphysics model of a moving mirror.

\section{Mirror-Oscillator-Field (MOF) model and quantum Brownian motion}
\label{sec:qbm}

In this Section we shall establish a connection between the MOF model for $N$ moving mirrors and $N$ harmonic oscillators interacting with a bath of harmonic oscillators that constitute an environment for the $N$ oscillators. The latter system has a long and well-developed history for providing a simple model with which to study quantum Brownian motion (QBM). Hence, if a relationship between the MOF model and QBM exists then one should be able to exploit the results of many previous studies (regarding decoherence, (dis)entanglement, fluctuation-dissipation relations, etc.) to apply towards moving mirror systems. We show here that such a relationship does indeed exist.

\subsection{Static mirrors and QBM}

Consider a mirror at rest that is fixed at $Z(t)=0$ for all time. The MOF Hamiltonian for this configuration follows from (\ref{nonrelH1})
\begin{align}
	H = \frac{1}{2} \int dx \big( \Pi^2 (x^\alpha) + (\partial_x \Phi (x^\alpha))^2 \big) + \frac{ p^2 (t) }{2m} + \frac{1}{2} m \Omega^2 q^2 (t) - \lambda q(t) \Phi(t, 0)  .
\end{align}
It is well known that a field can be represented as a continuum of harmonic oscillators, some of which have arbitrarily large natural frequencies. However, such large frequencies are not usually physically relevant (and often lead to divergences that must be properly handled with well-established renormalization techniques) so that one can simply impose a cut-off frequency $\Lambda$, which has the effect of ensuring that all calculated quantities are finite \cite{footnote3}.

The mode decomposition of the field is
\begin{align}
	\Phi (t, x) = \sum_k \sum_{\sigma=1}^2 \varphi_k ^\sigma (t) u_k ^\sigma (x)  .
\end{align}
If we restrict the field to the interior of a 1-dimensional (but large) volume $V$ then the normal modes of the field are simply
\begin{align}
	u_k^1 (x) & = ( 2 V \omega_k )^{-1/2} \cos k x \\
	u_k^2 (x) & = ( 2 V \omega_k )^{-1/2} \sin k x
\end{align}
so that the time dependence of the $k^{\rm th}$ mode has the following representation in terms of creation and annihilation operators
\begin{align}
	\varphi_k ^1 (t) & = a_k e^{- i \omega_k t} + a_k^\dagger e^{i \omega_k t} \\
	\varphi_k ^2 (t) & = i \big( a_k e^{- i \omega_k t} - a_k^\dagger e^{i \omega_k t} \big)
\end{align}
In terms of this mode decomposition, the Hamiltonian is
\begin{align}
	H = \frac{1}{2} \sum_k \sum_{\sigma=1}^2 \big( (\pi_k^\sigma )^2 + k^2 (\varphi_k^\sigma)^2 \big) + \frac{ p^2 }{ 2m } + \frac{1}{2} m \Omega^2 q^2 - \lambda \sum_k \sum_{\sigma=1}^2  q(t) u_k^\sigma (0) \, \varphi_k ^\sigma (t)  .
\label{eq:qbmH0}
\end{align}
Notice that the coupling constant $\lambda$ in the last term can be grouped with the mode function $u_k^\sigma(0)$ to give an effective coupling constant that depends on the particular mode $C_k^\sigma \equiv \lambda u_k ^\sigma (0)$. Therefore, the Hamiltonian for this system is
\begin{align}
	H = \frac{1}{2} \sum_k \sum_{\sigma=1}^2 \big( (\pi_k^\sigma )^2 + k^2 (\varphi_k^\sigma)^2 \big) + \frac{ p^2 }{ 2m } + \frac{1}{2} m \Omega^2 q^2 -  \sum_k \sum_{\sigma=1}^2  C_k^\sigma q(t)  \varphi_k ^\sigma (t) = H_{1{\rm -HO~QBM}} ,
\label{eq:qbmH1}
\end{align}
which is precisely the Hamiltonian for a harmonic oscillator $q(t)$ coupled to an environment composed of a bath of harmonic oscillators $\{ \varphi _k ^\sigma (t) \}$. In other words, the MOF model for a mirror at rest can be related to quantum Brownian motion where the field provides the environment that the mirosc interacts with. QBM has a long history and is well-studied so that results already found in that literature can be applied directly to the interaction of a field with a static mirror via the MOF model. For example, the master equation is exactly known for this system \cite{HPZ} and so one can study its behavior  near the perfectly-reflecting limit where $\lambda \to \infty$ or, equivalently, $m\to 0$ as well as in a non-zero temperature regime.

A similar result holds for $N$ mirrors held at rest at positions $x = L_a$ with $a = 1, \ldots, N$. It is straightforward to see that the corresponding Hamiltonian, when decomposing the field into harmonic oscillators, is
\begin{align}
	H = \frac{1}{2} \sum_k \sum_{\sigma=1}^2 \big( (\pi_k^\sigma )^2 + k^2 (\varphi_k^\sigma)^2 \big) + \sum_{a=1}^N \bigg( \frac{ p_a^2 }{ 2m_a } + \frac{1}{2} m_a \Omega_a^2 q_a^2 -  \sum_k \sum_{\sigma=1}^2  C_{k a}^\sigma q_a(t)  \varphi_k ^\sigma (t) \bigg) = H_{N  {\rm -HO~QBM}}
\label{eq:qbmH2}
\end{align}
where the effective bilinear coupling constant is $C_{ka}^\sigma \equiv \lambda u_k^\sigma (L_a)$. Therefore, $N$ static mirrors in the MOF model correspond to $N$ harmonic oscillators (mirosc variables) coupled to a bath of oscillators (the field). For $N=2$ oscillators coupled to a general environment, the exact master equation has been derived in \cite{ChouYuHu} and thus can be used to provide a different perspective and new insights in the description of a field coupled to two partially transmitting mirrors via the MOF model.

\subsection{Slowly moving mirrors and QBM}

Turn next to find the relationship between slowly moving mirrors in the MOF model and quantum Brownian motion. Let us first consider one mirror since the result for $N$ mirrors will generalize in an obvious way. Assume that the mirror is in an externally generated potential $V(x)$, such as a harmonic trap. Then the Hamiltonian in (\ref{eq:qbmH0}) is
\begin{align}
	H = \frac{1}{2} \sum_k \sum_{\sigma=1}^2 \big( (\pi_k^\sigma )^2 + k^2 (\varphi_k^\sigma)^2 \big) + \frac{ p^2 }{ 2m } + \frac{1}{2} m \Omega^2 q^2 + \frac{P^2}{2M } + V(Z) - \lambda \sum_k \sum_{\sigma=1}^2  q(t) u_k^\sigma (Z(t)) \, \varphi_k ^\sigma (t)
\label{eq:qbmH3}
\end{align}
where we have included the worldline variable to the Hamiltonian. Notice that from a QBM perspective, the effective coupling constant acquires a time dependence since the mode function is now time dependent, $u_k^\sigma ( Z(t))$. However, if the potential $V(x)$ restricts the motion of the mirror to be only small perturbations from its equilibrium position at $x=0$ then we may expand the mode function about the origin so that the interaction term above becomes
\begin{align}
	- \lambda \sum_k \sum_{\sigma=1}^2  q(t) u_k^\sigma (Z(t)) \, \varphi_k ^\sigma (t) = - \sum_k \sum_{\sigma=1}^2 C_k^\sigma q(t) \varphi_k^\sigma (t) - \lambda Z(t) \sum_k \sum_{\sigma=1}^2  \partial_x u_k^\sigma(0) q(t) \varphi_k^\sigma(t) + O(Z^2)
\end{align}
Therefore, the Hamiltonian (\ref{eq:qbmH3}) is equal to an unperturbed Hamiltonian, given by the $1$-harmonic oscillator QBM Hamiltonian in (\ref{eq:qbmH1}), plus an interaction Hamiltonian that describes perturbations due to the small displacement of the mirror that arise from interactions between the field oscillators and the mirosc,
\begin{align}
	H = H_{1 {\rm -HO~QBM}} - \lambda Z(t) \sum_k \sum_{\sigma=1}^2  \partial_x u_k^\sigma (0)q(t)  \varphi_k^\sigma(t) + O(Z^2)
\label{eq:qbmH4}
\end{align}
Hence, one can compute the perturbations of, for example, the exact master equation for 1-harmonic oscillator QBM to study the behavior of a movable, partially transmitting mirror. Notice that if $V(Z) = M \Omega_0^2 Z^2(t) / 2$ then (\ref{eq:qbmH4}) describes a nonlinearly coupled QBM system where the mirosc and the mirror's position are the two oscillators in an open system that couples to the bath provided by the field oscillators. The nonlinearity is only in the mirror's position (i.e., from the $O(Z^2)$ terms above) but the mirosc and the field oscillators still couple to each other bilinearly.

The generalization to $N$ mirrors should be obvious with the Hamiltonian describing the system being
\begin{align}
	H = H_{N {\rm -HO~QBM}} - \lambda \sum_{a=1}^N Z_a(t) \sum_k \sum_{\sigma=1}^2 \partial_x u_k^\sigma (L_a) q(t) \varphi_k^\sigma (t) + O(Z_a^2)
\end{align}
where the unperturbed position of the $a^{\rm th}$ mirror is at $x=L_a$. In particular, one can compute the perturbations of, for example, the exact master equation for 2-harmonic oscillator QBM \cite{ChouYuHu} to study  entanglement, decoherence, etc., of a cavity with movable, partially transmitting mirrors.

\section{Summary and Further Developments}
\label{sec:summary}

In this paper we constructed a microphysics model of moving mirrors interacting with a quantum field. The novel ingredient we introduced is a harmonic oscillator (a ``mirosc'') model describing the internal degrees of freedom of the mirror that couples to the incident radiation thereby providing a mechanism for the dynamical interplay of the mirror-field system. Since the field can transfer (receive) energy and momentum to (from) the mirosc the collection of them serves the function of a partially reflecting or transmitting mirror.  We showed that this mirror-oscillator-field (MOF) system can perfectly reflect or perfectly transmit radiation depending on the values of the mirosc mass $m$, natural frequency $\Omega$, and coupling strength $\lambda$ to the field. Perfect reflection can be attained in three ways: 1) $m \to 0$; 2) $\lambda \to \infty$; and 3) Frequency $\omega$ of an incident wave is equal to the mirosc natural frequency $\Omega$. Limits 1) and 2) exhibit perfect reflection (or nearly so) among a broad frequency bandwidth whereas limit 3) strongly reflects modes with frequencies near $\Omega$ because of a resonant excitation of the mirosc.

The MOF model reduces to several commonly used models of moving mirrors in a quantum field. We showed that when the mirosc variable $q(t)$ evolves adiabatically ($| \ddot{q} | \ll | \Omega^2 q|$) or when $m\to0$ but $m\Omega^2 = \kappa$ remains constant then the MOF model reduces to the Barton and Calogeracos (BC) model \cite{BC} of a partially transmitting moving mirror. The free parameter in the BC model $\gamma$ is related to the mirosc parameters of the MOF model $(m, \Omega, \lambda)$ by $\gamma = \lambda^2 / (2 m \Omega^2)$. The ``auxiliary field'' model of Golestanian and Kardar \cite{Kardar, KardarRMP} arises from the MOF model in the limit that $m\to 0$. In this limit, there is no mirosc dynamics and $q(t)$ becomes an auxiliary variable. In the quantum theory, $q(t)$ may have any possible realization (see (\ref{eq:Zmassless1})), which manifests as a Dirichlet boundary condition on the field at the location of the mirror and thus perfectly reflects incident radiation (see (\ref{eq:Zmassless2})). We also showed that our MOF model reduces to the phenomenological model of a mirror interacting with a cavity field via the radiation pressure exerted on the mirror's surface when a number of assumptions are made (though these may not all be  necessary to derive the $Nx$-coupling in other setups). This ``$Nx$-coupling'' is often used to describe laboratory setups but may be extended by the MOF model to scenarios where the mirosc does not evolve adiabatically, which may exhibit interesting macroscopic (or perhaps mesoscopic) quantum phenomena.
Additionally, $Nx$-type coupling provides the leading order corrections to the classical radiation pressure coupling when the cavity is occupied by low photon numbers. The model we present in this paper will remain useful even when the necessary conditions for it to match with models with $Nx$-type coupling are not met, for example, when the mirror motion is sufficient to excite field quant to higher modes.

The bulk motion of the mirror in the MOF model, which may be relativistic depending on the application, can be derived from an action or a Hamiltonian. In either formulation, we find that the mirror moves with a time-dependent effective mass $M_{\rm eff}$ that is composed of the mirror's rest mass $M$ and the mirror's total internal energy, which comes from the energy of the mirosc itself and its interaction with the field. We also showed (in a purely classical setting) that the MOF model seems to admit physical solutions despite the use of a point particle description for the mirror's motion and despite the interaction between the mirror and field resembling that of a charged particle (which can be plagued by pathologies). We demonstrated that when the field is generated by its interaction with the mirosc {\it alone} so that there is no external source $J_{\rm ext}$ and no initial field configuration present then the mirror will evolve on an inertial trajectory (i.e., constant velocity), which is the correct expected result, in contradistinction to the radiation reaction on a point charge in electrodynamics where the charge may exhibit run-away motions in the absence of any external forces acting on the charge.

As an application of the MOF model, we studied the``cooling'' of a mirror by its interactions with an external field in a purely classical context. We found that when the mirosc is weakly coupled to the field that the mirror, when perturbed, will oscillate around its equilibrium configuration while its displacement amplitude decays slowly in time.

An interesting consequence of our  MOF model for moving mirrors is that it relates to models of quantum Brownian motion (QBM) in a straightforward manner. The relation essentially follows because the field can be regarded as a continuum of harmonic oscillators. Hence, for $N$ mirrors held at rest, the MOF model is equivalent to $N$ harmonic oscillators in a bath of oscillators (from the field). For $N=1, 2$, the master equation for such a system in a general environment has been derived {\it exactly} \cite{HPZ,ChouYuHu} and even for general $N$ \cite{FlemingNHO}. Consequently, the MOF model can be used to study the superposition of two mirrors,  the decoherence by and the disentanglement of moving mirrors via a field, etc., so as to gain insight into these aspects of macroscopic quantum phenomena.  
We expect that the rich repository of technical tools and physical insights  from the study of QBM can be carried over directly to our MOF model for a broad range of  applications involving moving mirrors and quantum fields. For example, QBM results for systems at finite temperature may provide a simple way to incorporate thermal effects into the MOF model. We will begin to explore this theme in a follow-up paper \cite{StoEqOM} on the theory of OM from an open quantum system viewpoint.

The generalization of the MOF model to 3 spatial dimensions can be made where the mirror is an extended body having some surface geometry. On this surface, we may place a layer of mirosc's that play  the role of the electrons in a metal gas or dielectric medium providing the mirror's light degrees of freedom and responsible for reflection of incident radiation over some bandwidth of the electromagnetic spectrum (e.g., optical as in many metals). Incorporating the electromagnetic field in the MOF model should also be straightforward as its structure is similar to that of a minimally-coupled scalar field in the MOF model (see the appendix of \cite{RHA}).

In the second series on back-action effects we will study the full quantum mechanical evolution of the MOF system in the context of mirror cooling.
Therein, we will derive the exact equations of motion describing the mirror's average position. In the most general case we will show that the mirror motion is described by an integro-differential equation exhibiting non-Markovian dynamics. The equations can be simplified through a series of approximations which directly relate to experimentally engineerable quantities, such as the cavity's quality factor, and the relevant timescales for the mirror's internal dynamics. Given the broad range of applicability, these results can be employed to guide theoretical and experimental investigations ranging from the cooling of the center of mass motion of moveable mirrors, having broad-band reflective properties, to the manipulation of trapped ions near surfaces, possessing narrow-band reflective properties \cite{Hetet11}.

In the third series we will address the moving mirror analog of the back-reaction of Hawking radiation  \cite{Haw75} on the evolution of a black hole. There are controversies in some deep issues  related to the end-state of black hole evaporation resulting from the Hawking effect, namely, whether complete evaporation of a black hole means the non-unitary evolution of quantum states (see, e.g., \cite{Preskill}) which violates the basic tenets of theoretical physics or if unitarity is preserved, and if so, how?  One key ingredient, the back-reaction of the emitted radiation on the spacetime, has not been taken into account fully or correctly (for a recent update, see \cite{HuRou07,QFTextProc} and papers cited therein.) There are analog  studies on how information is shared in the black hole (harmonic) atom - quantum field system (see, e.g., \cite{LinHu08} and references therein.) as well as moving mirror analog problem  \cite{FulDav}. The connection was made between the $s$-wave component of Hawking evaporation and the emission of radiation from moving mirrors by the dynamical Casimir effect but, like the original calculation by Hawking, treated the effects of back reaction rather coarsely.  Since the MOF model offers a large degree of flexibility and tractability, we were able to find exact equations of motion for the mirror incorporating the effects of back-reaction \cite{BGH_bhbkrn}. These exact solutions, as well as those from the atom-field analogs, can provide new insights into this basic issue in theoretical physics.

\section{acknowledgements}

CG was supported in part by an appointment to the NASA Postdoctoral Program at the Jet Propulsion Laboratory administered by Oak Ridge Associated Universities through a contract with NASA, and  in part by a NIST Gaithersburg grant awarded to the University of Maryland when this work was started. 
RB gratefully acknowledges the support of the U.S. Department of Energy through the LANL LDRD program.
BLH wishes to thank Professor Jason Twamley, Director of the the Centre for Quantum Computer Technology at Macquarie University for his warm hospitality in Feb-Mar 2011 during which this work was partly carried out. His research was partially supported by NSF grant PHY-0801368 to the University of Maryland. Copyright 2012. All rights reserved.



\begin{thebibliography}{99}
\bibitem{footnote1}
An analogous endeavor undertaken by one of us and many others in the early 70s is quantum field theory (QFT) in curved spacetime \cite{BirDav}, which is the theory behind cosmological particle creation and Hawking-Unruh effects. It contains as a subset, QFT with boundaries, which deals with effects due to geometry, topology, such as the Casimir effect \cite{Casimir} and due to finite size of the system \cite{OSH}. In its latest reincarnation this is  the subject of QFT under external conditions, see, e.g., \cite{QFTextProc}
\bibitem{footnote}
Actually, to describe the nonequilibrium dynamics of this quantum theory one should use an initial value formulation of the path integral. Such a prescription is given by the ``in-in'' or ``closed-time-path'' framework (see e.g., \cite{Schwinger:JMathPhys2, HPZ, CalHu08} and citing references) and has been used for moving mirror problems previously by \cite{FLM}.
\bibitem{footnote2}
The homogeneous solution for the field that we are neglecting here would play the role of a Langevin forcing term when plugged into the remaining equations of motion, and would account for the initial state of the field. In a future paper this term will be considered and will account for the field's thermal and quantum fluctuations.
\end{thebibliography}
\end{document}